\documentclass[onecolumn,english,aps,showpacs,preprintnumbers,amsmath,amssymb,floatfix]{revtex4-1}

\usepackage{graphicx}
\usepackage{float}
\usepackage[colorlinks=true,linkcolor=blue,urlcolor=blue,citecolor=blue]{hyperref}
\usepackage[mathlines, pagewise]{lineno}\linenumbers\relax \modulolinenumbers[1]

\usepackage{soul}
\begin{document}
	\title{Laplace-Space Analysis of $xF_3$ Including Nuclear Effects and Gegenbauer-Polynomial Parton Distributions }

	\author{Shahin Atashbar Tehrani$^{1,\href{https://orcid.org/0000-0002-9279-499X}{\textsuperscript{\textcolor{green}{\Large$\circ$}}}}$}
		\email{Atashbart3@gmail.com}

		\author{Javad Sheibani$^{2,\href{https://orcid.org/0000-0002-9967-7059}{\textsuperscript{\textcolor{green}{\Large$\circ$}}}}$}
	\email{J.sheibani@ipm.ir}

	\author{Elham Astaraki$^{3,\href{https://orcid.org/0009-0007-8216-3389}{\textsuperscript{\textcolor{green}{\Large$\circ$}}}}$}
	\email{Astaraki.elham@razi.ac.ir}
	
	\author{Reyhaneh Farashaeian$^{2,5,\href{https://orcid.org/0009-0000-9573-0273}{\textsuperscript{\textcolor{green}{\Large$\circ$}}}}$}
	\email{farashaeian@ipm.ir}
	
		\author{Kourosh (Rouhollah) Gheisari$^{4,\href{https://orcid.org/0000-0001-5312-3811}{\textsuperscript{\textcolor{green}{\Large$\circ$}}}}$}
	\email{gheisari@pgu.ac.ir}
		
	\affiliation{ 		
		$^1$	Department of Physics, Faculty of Nano and Bio Science and Technology, Persian Gulf University, 75169 Bushehr, Iran\\	
			$^2$	School of Particles and Accelerators, Institute for Research in Fundamental Sciences (IPM), P.O.Box 19395-5531, Tehran, Iran.\\
		$^{(3)}$Department of Physics, Razi University, Kermanshah 67149, Iran\\
		$^{(4)}$Physics Department,Yasouj University,75918-74934
		,Yasouj,Iran\\
		$^{(5)}$Physics Department,Shahid Beheshti University,Tehran,Iran
	}%
	
	\date{\today}
\begin{abstract}
	We present a next-to-leading-order (NLO) and next-to-next-to-leading-order (NNLO) QCD analysis of the structure functions $xF_3$, $F_2^p$, $F_2^d$, and the non-singlet combination $F_2^{\rm NS}$, with the primary aim of extracting the valence-quark parton distribution functions (PDFs) from the available experimental data. The input PDFs are parametrized using a flexible Gegenbauer-polynomial expansion. We solve the DGLAP evolution equations analytically in Laplace space and reconstruct the structure functions in Bjorken-$x$ space using a Jacobi-polynomial expansion. Nuclear correction factors are included in the analysis of the $xF_3$ neutrino--nucleus data, whereas no nuclear corrections are applied to the deuteron structure function $F_2^d$. We use the measured structure functions to constrain the valence-quark distributions and examine the effects of the perturbative order and the treatment of nuclear corrections on their extraction. The resulting analysis provides a consistent framework for determining valence-quark PDFs from proton, deuteron, and neutrino--nucleus data over the kinematic range considered.
\end{abstract}


 \maketitle

\section{Introduction}

The analysis uses experimental data for the structure functions $xF_3$, $F_2^p$, $F_2^d$, and $F_2^{\rm NS}$ to constrain the valence-quark distributions. The $xF_3$ dataset comprises 230 data points from the CCFR~\cite{Seligman:1997mc}, NuTeV~\cite{NuTeV:2005wsg}, and CHORUS~\cite{CHORUS:2005cpn} experiments, with 116, 64, and 50 points, respectively; all 230 points are retained after the applied cuts. The proton structure-function dataset, $F_2^p$, includes 467 points from BCDMS~\cite{Benvenuti:1989fm,Benvenuti:1989rh,Benvenuti:1989gs}, NMC~\cite{Arneodo:1996qe,Arneodo:1995cq}, SLAC~\cite{Whitlow:1991uw}, H1~\cite{H1}, and ZEUS~\cite{ZEUS}, of which 248 remain after the cuts. The deuteron dataset, $F_2^d$, contains 232 points from BCDMS, NMC, and SLAC, with 159 points retained after the cuts. No nuclear corrections are applied to $F_2^d$ in this analysis. The non-singlet dataset, $F_2^{\rm NS}$, consists of 208 points from BCDMS, NMC, and SLAC, of which 165 remain after the cuts. Nuclear correction factors are included in the treatment of the neutrino--nucleus data used for $xF_3$. The data from all four structure functions are analyzed consistently to extract the valence-quark distributions.

At the heart of any QCD analysis lies the Dokshitzer-Gribov-Lipatov-Altarelli-Parisi (DGLAP) evolution equations~\cite{evol1,evol2,evol3,evol4}, which describe how parton distributions evolve with the momentum transfer $Q^2$. Solving these integro-differential equations with sufficient precision across the full kinematic domain represents a significant computational undertaking. Traditional approaches rely on numerical integration techniques, employing discretized grids in Bjorken-$x$ space. While sophisticated packages such as QCDNUM~\cite{Botje:2010ay} and APFEL~\cite{Bertone:2013vaa} have proven remarkably successful, they are not without limitations. These methods require careful optimization of grid parameters, can become computationally expensive when performing repeated fits with varying theoretical assumptions, and propagate parametric uncertainties in ways that are sometimes difficult to track systematically. Furthermore, the finite grid resolution can introduce numerical artifacts, particularly in regions where the PDFs exhibit steep behavior, such as the small-$x$ and large-$x$ domains. These considerations have motivated the search for alternative formulations that offer greater analytical control and computational efficiency.

In recent years, spectral methods and integral transforms have emerged as powerful alternatives to purely numerical approaches for solving the DGLAP equations. These techniques offer several compelling advantages: they provide analytical or semi-analytical solutions, reduce the complexity of the evolution equations from integro-differential to ordinary differential form, and naturally facilitate the propagation of uncertainties through the fit procedure. A particularly elegant implementation of this philosophy was developed in Ref.~\cite{MoosaviNejad:2016ebo,Khanpour:2016uxh},A combined Laplace--Jacobi framework was employed at next-to-leading order (NLO) and next-to-next-to-leading order (NNLO) in perturbative QCD to analyze the structure functions $xF_3$, $F_2^p$, $F_2^d$, and $F_2^{\rm NS}$. In this framework, the DGLAP evolution equations are treated in Laplace space, while the Jacobi-polynomial expansion is used to reconstruct all four structure functions in Bjorken-$x$ space.

Recent studies have investigated valence-quark distributions using complementary experimental and methodological approaches. These analyses have simultaneously incorporated proton and deuteron structure-function data, $F_2^p$ and $F_2^D$, the nonsinglet structure function $F_2^{\mathrm{NS}}$, and the structure function $xF_3$~\cite{Shahrzad:2024ucw}. In addition, neural-network techniques~\cite{Astaraki:2026wnh} have been employed to extract the valence distributions with increased flexibility and reduced dependence on a fixed functional form.

The essence of this methodology lies in a transformation from Bjorken-$x$ space to the logarithmic variable $\nu = \ln(1/x)$, followed by the application of a Laplace transform. This mapping effectively converts the convolution integrals that characterize the DGLAP evolution into ordinary products in Laplace space. The resulting equations become decoupled first-order ordinary differential equations in the evolution variable $\tau = \ln(Q^2/\Lambda^2)$, which can be solved analytically. The evolved Laplace-space distributions, when multiplied by the appropriate NLO and NNLO Wilson coefficient functions, yield the transformed structure function $F_3(s,\tau)$. To reconstruct the physical $x$-dependent structure function, a Jacobi polynomial expansion is employed~\cite{Kataev:2001kk,Kataev:1998ce,parisi}:

\begin{equation}
xF_3(x,Q^2) = x^{\beta}(1-x)^{\alpha}\sum_{n=0}^{N_{\max}} \Theta_n^{\alpha,\beta}(x)\, a_n(Q^2),
\end{equation}

where the expansion coefficients $a_n(Q^2)$ are constructed directly from a finite set of Laplace moments, typically with $N_{\max} \approx 9$. This formalism achieves remarkable numerical accuracy while maintaining computational efficiency, as the analytical solution eliminates the need for iterative numerical integration.

While Refs.~\cite{MoosaviNejad:2016ebo,Khanpour:2016uxh} demonstrated the effectiveness of the Laplace--Jacobi framework in describing the CCFR, NuTeV, and CHORUS neutrino--nucleus data for $xF_3$, the present work extends the analysis to a global fit of the $xF_3$, $F_2^p$, $F_2^d$, and $F_2^{\rm NS}$ datasets, with the objective of extracting the valence-quark distributions. The input valence distributions, $xu_v$ and $xd_v$, are described using a flexible Gegenbauer-polynomial parameterization, and all four structure functions are analyzed within the same Laplace--Jacobi framework at NLO and NNLO accuracy. Nuclear correction factors are incorporated in the treatment of the $xF_3$ neutrino--nucleus data, while no nuclear corrections are applied to the proton, deuteron, or non-singlet datasets. This combined fit enables the different datasets to constrain the valence-quark distributions simultaneously, while keeping the treatment of nuclear effects specific to the $xF_3$ data.

The present work is motivated by the desire to address both of these limitations within a unified framework. To this end, we introduce two key extensions to the Laplace-Jacobi formalism. First, we adopt a highly flexible parameterization of the valence PDFs based on Gegenbauer polynomials. These polynomials form a complete orthogonal set on the interval $x\in[0,1]$ with a suitable weight function, and offer superior convergence properties compared to conventional parameterizations. The Gegenbauer expansion provides a systematic and controllable means of increasing the flexibility of the input distributions without introducing unphysical oscillations, making it particularly well-suited for capturing the non-perturbative dynamics of valence quarks. Second, we dynamically incorporate nuclear modification factors directly into the input PDFs at the initial scale $Q_0^2$. This is achieved by defining the nuclear-modified distributions as $q_v^{A}(x,Q_0^2) = \mathcal{W}_{q_v}(x,A,Z,N)\, q_v(x,Q_0^2)$, and using these as the initial conditions for the Laplace-space DGLAP evolution. This formulation ensures that the nuclear corrections are propagated consistently through the evolution equations, thereby capturing their full $Q^2$ dependence rather than treating them as scale-independent modifications.

The analytical solution of the DGLAP equations in Laplace space has a rich history, with foundational contributions establishing the basic framework for treating parton evolution in transform space~\cite{Block:2010ti,Block:2009en}. Subsequent developments, particularly the implementation of Jacobi polynomials, enabled the precise reconstruction of $x$-dependent distributions from their Laplace and Mellin moments~\cite{AtashbarTehrani:2013qea,Nematollahi:2023dvj,MoosaviNejad:2019ybp,Sheibani:2018gxt,Taghavi-Shahri:2016ktz,Khanpour:2016uxh,Zareei:2015gkg}. These analytical methods have been continually refined and applied to global QCD fits, most notably in the Laplace-Jacobi analysis of Ref.~\cite{MoosaviNejad:2016ebo}, which provided a comprehensive description of $xF_3$ at NLO and NNLO.

With the enhanced Gegenbauer--Laplace framework developed in this work, we perform a comprehensive global analysis of the $xF_3$, $F_2^p$, $F_2^d$, and $F_2^{\rm NS}$ data from the CCFR, NuTeV, CHORUS, BCDMS, NMC, SLAC, H1, and ZEUS experiments.
 In addition to extracting the valence distributions, we investigate the implications for three fundamental sum rules: the Gross-Llewellyn Smith, Bjorken, and Adler sum rules. These sum rules, derived from current algebra and symmetry principles, provide stringent tests of the internal consistency of our analysis and serve as sensitive probes of nuclear corrections. By comparing the nuclear-modified predictions with both the free-proton values and the exact theoretical expectations, we quantify the magnitude of nuclear effects and assess their impact on precision tests of QCD. The paper is organized as follows: Section~\ref{sec2} presents the theoretical framework, including the Gegenbauer parameterization and the nuclear modification factors. Section \ref{sec3} describes evolution in Laplace space and in the Sec. \ref{sec4} global fit methodology and the experimental data sets presents our results for the valence PDFs, structure functions,In Section \ref{sec5} we discus about sum rules. Finally, Section~\ref{sec6} contains our conclusions and outlook.

\section{Theoretical Framework}\label{sec2}

Parton distribution functions provide the essential bridge between the non-perturbative structure of hadrons and the perturbative calculations of high-energy scattering processes. Within the proton, these functions describe the momentum fraction $x$ carried by each flavor of quark or gluon. For charged-current neutrino scattering, the structure function $xF_3$ is particularly valuable because it isolates the valence quark content of the nucleon. The valence distributions $xu_v$ and $xd_v$ represent the net contributions of up and down quarks, carrying the baryon number and satisfying the fundamental sum rules $\int_0^1 u_v\,dx = 2$ and $\int_0^1 d_v\,dx = 1$. Unlike sea quarks and gluons, these distributions are directly constrained by the quantum numbers of the proton and dominate the behavior of $xF_3$ across most of the kinematic range.

At the initial scale $Q_0^2 = 1\,\mathrm{GeV}^2$, we parameterize the valence distributions using an expansion in Gegenbauer polynomials, following the approach introduced in Ref.~\cite{Arbabifar:2024wga}. The Gegenbauer expansion provides systematic convergence, allowing flexible functional forms while avoiding the spurious oscillations that can plague high-order polynomial fits. The up-valence distribution is expressed as:

\begin{equation}\label{eq:xuvQ0}
xu_v(x,Q_0^2) = \mathcal{N}_u \, x^{\alpha_{u_v}} (1-x)^{\beta_{u_v}} 
\left(1 + \sum_{i=1}^{3} a_u^i \, C_i^{7/2}(1-2x)\right),
\end{equation}

where $C_i^{7/2}$ denotes the Gegenbauer polynomial of order $i$ with parameter $7/2$. The prefactor $x^{\alpha_{u_v}}$ governs the behavior as $x\to 0$, while $(1-x)^{\beta_{u_v}}$ controls the regime $x\to 1$. The polynomial sum adds flexibility in the intermediate region, where experimental neutrino data provide the strongest constraints. For the down-valence distribution, we adopt a parameterization that shares the same Gegenbauer coefficients as the up sector while allowing independent behavior near $x=1$:

\begin{eqnarray} \label{eq:xdvQ0}
xd_v(x,Q_0^2) &=& \frac{\mathcal{N}_d}{\mathcal{N}_u} (1-x)^{\beta_{d_v}} xu_v(x,Q_0^2) \nonumber\\
&=& \mathcal{N}_d \, x^{\alpha_{u_v}} (1-x)^{\beta_{u_v}+\beta_{d_v}} 
\left(1 + \sum_{i=1}^{3} a_u^i \, C_i^{7/2}(1-2x)\right).
\end{eqnarray}
\begin{figure}[h]
	\begin{center}
		\includegraphics[clip,width=0.5\textwidth]{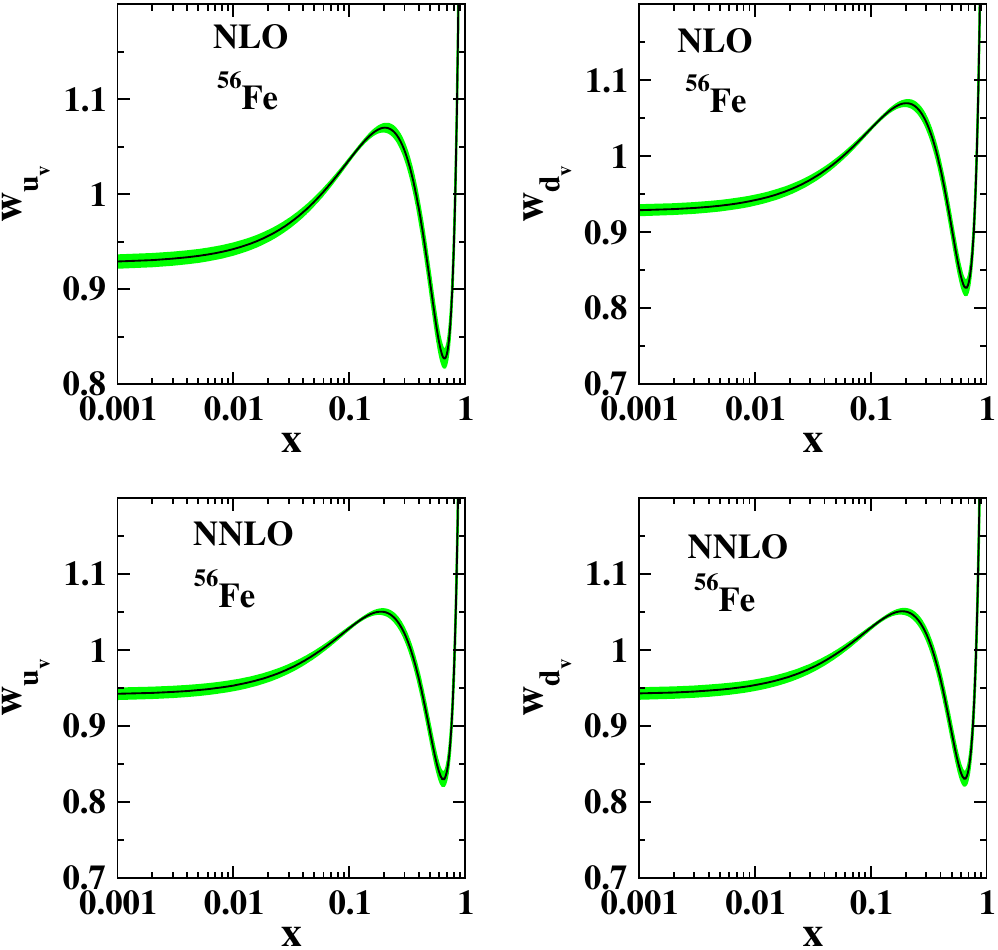}%
		\caption{\sf The nuclear weight function calculated in refs.~\cite{AtashbarTehrani:2012xh,Khanpour:2016pph} in NLO and NNLO approximation for $^{56}Fe$. } \label{fig1}
	\end{center}
\end{figure}
\begin{figure}[h]
	\begin{center}
		\includegraphics[clip,width=0.5\textwidth]{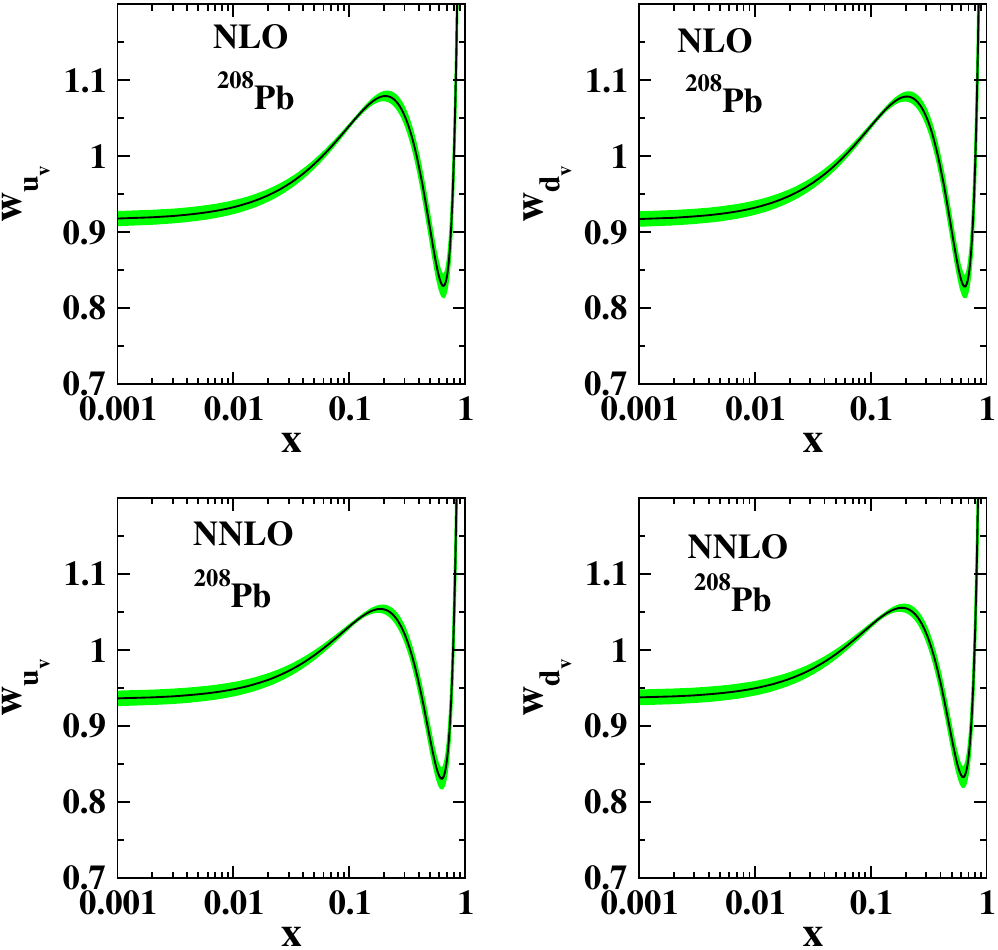}%
		\caption{\sf The nuclear weight function calculated in the refs.~\cite{AtashbarTehrani:2012xh,Khanpour:2016pph} in NLO and NNLO approximation for $^{208}Pb$. } \label{fig2}
	\end{center}
\end{figure}
The factor $(1-x)^{\beta_{d_v}}$ introduces a different large-$x$ fall-off for the down quarks compared to the up quarks, reflecting the fact that the proton contains two up quarks and one down quark, leading to distinct endpoint behavior. The normalization constants $\mathcal{N}_u$ and $\mathcal{N}_d$ are fixed through the valence sum rules, ensuring the correct number of valence quarks in the proton.

The experimental data employed in this analysis originate from neutrino scattering on nuclear targets: iron (${}^{56}\mathrm{Fe}$) from the CCFR and NuTeV collaborations, and lead (${}^{208}\mathrm{Pb}$) from CHORUS. Bound nucleons within a nucleus experience a modified environment compared to free nucleons, leading to alterations in their parton distributions. These nuclear effects arise from multiple physical sources: at small $x$, gluon shadowing and sea-quark shadowing suppress the distributions; in the intermediate region, anti-shadowing enhances them; and at large $x$, Fermi motion and binding corrections reduce the momentum fractions carried by quarks. The combined effect, known as the EMC effect, represents a substantial modification of the nuclear structure. To incorporate these phenomena consistently, we introduce nuclear weight functions $\mathcal{W}_{u_v}$ and $\mathcal{W}_{d_v}$ that relate the nuclear distributions $q_v^A$ to their free-nucleon counterparts.

For a nucleus with atomic number $Z$, neutron number $N$, and mass number $A=Z+N$, the valence distributions are constructed as:

\begin{eqnarray}
xu_v^A(x,Q_0^2) &=& \mathcal{W}_{u_v}(x) \, \frac{Z\, xu_v(x,Q_0^2) + N\, xd_v(x,Q_0^2)}{A}, \nonumber\\
xd_v^A(x,Q_0^2) &=& \mathcal{W}_{d_v}(x) \, \frac{N\, xu_v(x,Q_0^2) + Z\, xd_v(x,Q_0^2)}{A}.
\end{eqnarray}

In the limit $\mathcal{W}_{u_v} = \mathcal{W}_{d_v} = 1$, these expressions reduce to the isoscalar average of proton and neutron distributions weighted by the appropriate nucleon numbers, as expected for a system of non-interacting nucleons. The weight functions themselves are parameterized as:

\begin{eqnarray}
\mathcal{W}_{u_v}(x) &=& 1 + \left(1 - \frac{1}{A^{1/3}}\right) 
\frac{A_u + c_1 x + c_2 x^2 + c_3 x^3}{(1-x)^{0.4}}, \nonumber\\
\mathcal{W}_{d_v}(x) &=& 1 + \left(1 - \frac{1}{A^{1/3}}\right) 
\frac{A_d + c_1 x + c_2 x^2 + c_3 x^3}{(1-x)^{0.4}}.
\end{eqnarray}

The structure of this parameterization is designed to capture the essential physics of nuclear modifications across the entire $x$ range. The term $(1-x)^{-0.4}$ ensures that the effects of Fermi motion are properly represented as $x\to 1$. The polynomial in the numerator $A_{q} + c_1 x + c_2 x^2 + c_3 x^3$ provides the flexibility needed to describe shadowing at low $x$, anti-shadowing around $x\sim 0.1$, and the EMC suppression at large $x$. The prefactor $\left(1 - A^{-1/3}\right)$ scales with the nuclear size, reflecting the surface-to-volume ratio: as $A$ increases, the fraction of nucleons in the interior rises, and the magnitude of nuclear modifications grows accordingly. For $A=1$, the factor vanishes, recovering the free-nucleon limit.

The parameters $A_u$, $A_d$, $c_1$, $c_2$, and $c_3$ have been determined in Refs.~\cite{AtashbarTehrani:2012xh,Khanpour:2016pph} through global fits to nuclear DIS data at NLO and NNLO. The resulting weight functions for iron and lead are shown in Figs.~\ref{fig1} and~\ref{fig2}, respectively. These figures reveal the characteristic pattern of nuclear modifications: suppression below $x\simeq 0.1$, enhancement in the interval $0.1 \lesssim x \lesssim 0.3$, and suppression above $x\gtrsim 0.3$. The magnitude of these effects is larger for lead than for iron, consistent with the stronger nuclear binding in heavier systems. In our analysis, we adopt the weight functions from Refs.~\cite{AtashbarTehrani:2012xh,Khanpour:2016pph} and apply them at the initial scale $Q_0^2$. The resulting nuclear-modified distributions $q_v^A(x,Q_0^2)$ serve as the starting point for the DGLAP evolution, ensuring that the nuclear corrections evolve dynamically with $Q^2$. This treatment is superior to approaches that apply static corrections after evolution, since the scale dependence of nuclear effects is naturally incorporated. For iron, we use $A=56$, $Z=26$, $N=30$; for lead, $A=208$, $Z=82$, $N=126$. These values enter both the weighting factors in the nuclear valence distributions and the normalization of the weight functions through the $A^{-1/3}$ dependence.

\section{The Evolution in Laplace Space}\label{sec3}
In Laplace $s$-space, the valence-quark contribution to the proton
structure function ${\cal M}_2^p$ in the region $x \geq 0.3$ can be
written at NLO as
\begin{eqnarray}
	{\cal M}_2^p(s,\tau)
	&=&
	\left(\frac{4}{9}u_v(s)+\frac{1}{9}d_v(s)\right)
	\left(1+\frac{\tau}{4\pi}C_2^{(1)}(s)\right)
	e^{\tau\left[\Phi_{qq}^{LO}(s)+\frac{\tau_2}{\tau}	\Phi_{qq}^{NLO}(s)\right]} .
	\label{eq:FpNLO}
\end{eqnarray}
At NNLO, the corresponding expression becomes
\begin{eqnarray}
	{\cal M}_2^p(s,\tau)
	&=&
	\left(\frac{4}{9}u_v(s)+\frac{1}{9}d_v(s)\right)
	\left[
	1+\frac{\tau}{4\pi}C_2^{(1)}(s)
	+\left(\frac{\tau}{4\pi}\right)^2 C_2^{(2)}(s)
	\right]
	e^{\tau\left[\Phi_{qq}^{LO}(s)+\frac{\tau_2}{\tau}	\Phi_{qq}^{NLO}(s)
		+\frac{\tau_3}{\tau}\Phi_{\rm qq}^{NNLO}(s)\right]} .
	\label{eq:FpNNLO}
\end{eqnarray}

In the same kinematic region, the deuteron structure function is given
by
\begin{eqnarray}
	{\cal M}_2^d(s,\tau)
	&=&
	\frac{5}{18}\left[u_v(s)+d_v(s)\right]
	\left(1+\frac{\tau}{4\pi}C_2^{(1)}(s)\right)
	e^{\tau\left[\Phi_{qq}^{LO}(s)+\frac{\tau_2}{\tau}	\Phi_{qq}^{NLO}(s)\right]}
	\label{eq:FdNLO}
\end{eqnarray}
at NLO, and by
\begin{eqnarray}
	{\cal M}_2^d(s,\tau)
	&=&
	\frac{5}{18}\left[u_v(s)+d_v(s)\right]
	\left[
	1+\frac{\tau}{4\pi}C_2^{(1)}(s)
	+\left(\frac{\tau}{4\pi}\right)^2 C_2^{(2)}(s)
	\right]
	e^{\tau\left[\Phi_{qq}^{LO}(s)+\frac{\tau_2}{\tau}	\Phi_{qq}^{NLO}(s)
		+\frac{\tau_3}{\tau}\Phi_{\rm qq}^{NNLO}(s)\right]}
	\label{eq:FdNNLO}
\end{eqnarray}
at NNLO, where $d=(p+n)/2$.

For the difference between the proton and deuteron structure functions
in the region $x \leq 0.3$, we define
\begin{eqnarray}
	{\cal M}_2^{\rm NS}(s,\tau)
	&\equiv& 2\left[{\cal M}_2^p-{\cal M}_2^d\right](s,\tau)
	\nonumber\\
	&=&
	\left[
	\frac{1}{3}(u_v-d_v)(s)
	+\frac{2}{3}(\bar{u}-\bar{d})(s)
	\right]
	\left(1+\frac{\tau}{4\pi}C_2^{(1)}(s)\right)
	e^{\tau\left[\Phi_{qq}^{LO}(s)+\frac{\tau_2}{\tau}	\Phi_{qq}^{NLO}(s)\right]}
	\label{eq:FNSNLO}
\end{eqnarray}
at NLO. At NNLO, this becomes
\begin{eqnarray}
	{\cal M}_2^{\rm NS}(s,\tau)
	&=&
	\left[
	\frac{1}{3}(u_v-d_v)(s)
	+\frac{2}{3}(\bar{u}-\bar{d})(s)
	\right]
	\left[
	1+\frac{\tau}{4\pi}C_2^{(1)}(s)
	+\left(\frac{\tau}{4\pi}\right)^2 C_2^{(2)}(s)
	\right]
	e^{\tau\left[	\Phi_{qq}^{LO}(s)+\frac{\tau_2}{\tau}	\Phi_{qq}^{NLO}(s)
		+\frac{\tau_3}{\tau}\Phi_{\rm qq}^{NNLO}(s)\right]} .\nonumber\\
	\label{eq:FNSNNLO}
\end{eqnarray}
All explicit functional forms are provided in Appendix~\ref{AppendixA}.

Since sea-quark contributions cannot be neglected for $x\lesssim 0.3$,
we use the $\bar{u}-\bar{d}$ distribution from the CT18 parton
distribution functions~\cite{Hou:2019efy} at the input scale
$Q_0^2=1~\mathrm{GeV}^2$. The corresponding distribution is obtained as
\begin{equation}
	x(\bar{u}-\bar{d})(x,Q_0^2)
	=
	x\bar{u}(x,Q_0^2)-x\bar{d}(x,Q_0^2),
\end{equation}
where the CT18 PDFs are evaluated at $Q_0^2=1~\mathrm{GeV}^2$. To determine the \(x(\bar{u}-\bar{d})\) distribution for CT18, we used APFEL~\cite{Bertone:2013vaa} and LHAPDF~\cite{Buckley:2014ana} to extract it at the initial scale \(Q_0^2 = 1\,\mathrm{GeV}^2\). We then fitted an appropriate function to the extracted distribution.

It is worth noting here that the ${\text Q}^2$-dependence of the above evolution equations is expressed thorough the variable $\tau$ as $\tau (Q^2, Q_0^2) \equiv \frac{1}{4 \pi} \int_{Q_0^2}^{Q^2} \alpha_s ({Q^{\prime}}^{2}) d\ln {Q^{\prime}}^{2}$. and also:
\begin{equation}
	\tau_2 \equiv \frac{1}{(4 \pi)^2}
	\int_{Q_0^2}^{Q^2}
	\alpha_s^2(Q'^2)d \ln Q'^2\,,
\end{equation}
and
\begin{equation}
	\tau_3 \equiv \frac{1}{(4 \pi)^3}
	\int_{Q_0^2}^{Q^2} \alpha_s^3
	(Q'^2)d \ln Q'^2\,.
\end{equation}

The analytical solution of the DGLAP evolution equations in Laplace space provides an efficient framework for analyzing the structure functions included in our global fit. We consider the non-singlet structure function $xF_3$, together with the proton, deuteron, and non-singlet structure functions $F_2^p$, $F_2^d$, and $F_2^{\rm NS}$. The corresponding experimental data are taken from the CCFR, NuTeV, and CHORUS collaborations for $xF_3$; from BCDMS, NMC, SLAC, H1, and ZEUS for $F_2^p$; and from BCDMS, NMC, and SLAC for $F_2^d$ and $F_2^{\rm NS}$. These datasets are analyzed simultaneously in a global fit to determine the valence-quark distributions.

The method maps Bjorken-$x$ to the logarithmic variable $\nu=\ln(1/x)$ and applies a Laplace transform with respect to $\nu$. In Laplace space, the integro-differential DGLAP equations are converted into evolution equations in the scale variable $\tau=\ln(Q^2/\Lambda^2)$. For the non-singlet sector, which governs the evolution of the valence distributions and the non-singlet structure functions, the equations are diagonal in Laplace space and can be solved analytically. The evolved distributions are then used to calculate the structure functions at the desired values of $Q^2$.

For the $xF_3$ data, which are measured in neutrino--nucleus scattering, nuclear correction factors are incorporated into the analysis. No nuclear corrections are applied to the $F_2^p$, $F_2^d$, or $F_2^{\rm NS}$ datasets. Thus, the nuclear treatment is specific to the $xF_3$ data, while all four structure-function datasets contribute simultaneously to the global extraction of the valence-quark distributions.

The final reconstruction in Bjorken-$x$ space is performed using a Jacobi-polynomial expansion. For each structure function $F_i$, where
$i\in\{xF_3,F_2^p,F_2^d,F_2^{\rm NS}\}$, we write
\begin{equation}
	F_i(x,Q^2)
	=
	x^{\beta_i}(1-x)^{\alpha_i}
	\sum_{n=0}^{N_{\max}}
	\Theta_n^{\alpha_i,\beta_i}(x)\,
	a_{n,i}(Q^2),
	\label{eq:jacobi_expansion_all}
\end{equation}
where $\Theta_n^{\alpha_i,\beta_i}(x)$ are Jacobi polynomials and
$a_{n,i}(Q^2)$ are the corresponding expansion coefficients, obtained from the moments evolved in Laplace space. The same Laplace--Jacobi framework is therefore used to describe $xF_3$, $F_2^p$, $F_2^d$, and $F_2^{\rm NS}$ at both NLO and NNLO. Their combined description provides the basis for the global determination of the valence-quark PDFs.

In a precision QCD analysis, all corrections that can be comparable in
magnitude to the effects under study should be considered. In particular,
target-mass corrections (TMCs)~\cite{Georgi:1976ve}
are formally power-suppressed corrections to the leading-twist structure
functions. They can nevertheless be important at large Bjorken $x$ and
moderate $Q^2$, where the available constraints on the parton distribution
functions (PDFs) are also less stringent. An accurate treatment of TMCs is
therefore necessary when data in this kinematic region are included in a
perturbative QCD analysis.

We calculate the TMCs using the prescription of
Refs.~\cite{Georgi:1976ve,Gluck:2006yz,Steffens:2012jx,Khorramian:2009xz},
formulated in Laplace $s$-space. For a non-singlet structure function
${\cal M}_2^k$, the target-mass-corrected moment is written, retaining
terms through order $(m_N^2/Q^2)^2$, as
\begin{eqnarray}
	{\cal M}_{2,\mathrm{TMC}}^k(s,Q^2)
	&\equiv&
	{\cal L}\!\left[
	{\cal M}_{2,\mathrm{TMC}}^k(e^{-v},Q^2)
	\right](s)
	\nonumber\\
	&=&
	{\cal M}_2^k(s,Q^2)
	+\frac{s(s-1)}{s+2}
	\left(\frac{m_N^2}{Q^2}\right)
	{\cal M}_2^k(s+2,Q^2)
	\nonumber\\
	&&+
	\frac{\Gamma(s+3)^2}
	{2\Gamma(s-1)\Gamma(s+5)}
	\left(\frac{m_N^2}{Q^2}\right)^2
	{\cal M}_2^k(s+4,Q^2)
	+\mathcal{O}\!\left[
	\left(\frac{m_N^2}{Q^2}\right)^3
	\right].
	\label{equ:TMC}
\end{eqnarray}
Here, $m_N$ is the nucleon mass, and ${\cal M}_2^k(s,Q^2)$ denotes the
Laplace-space moment of the corresponding leading-twist non-singlet
structure function. In the following, we retain the terms displayed
explicitly in Eq.~(\ref{equ:TMC}) and neglect contributions of order
$(m_N^2/Q^2)^3$ and higher.

The target-mass-corrected structure function in Bjorken-$x$ space is then
reconstructed using the Jacobi-polynomial expansion:
\begin{eqnarray}
	F_2^{k,\mathrm{TMC}}(x,Q^2)
	&=&
	x^\beta(1-x)^\alpha
	\sum_{n=0}^{N_{\max}}
	\Theta_n^{\alpha,\beta}(x)
	\sum_{j=0}^{n}
	c_j^{(n)}(\alpha,\beta)\,
	{\cal M}_{2,\mathrm{TMC}}^k(j+1,Q^2).
	\label{eg1JacobTMC}
\end{eqnarray}
In this expression, ${\cal M}_{2,\mathrm{TMC}}^k(j+1,Q^2)$ is obtained
from Eq.~(\ref{equ:TMC}), while $\Theta_n^{\alpha,\beta}(x)$ and
$c_j^{(n)}(\alpha,\beta)$ are the Jacobi-polynomial basis functions and
their expansion coefficients, respectively.

In the present analysis, TMCs are included in the calculation of the
relevant structure functions using the prescription above. This treatment
is distinct from nuclear corrections: nuclear corrections are applied to
the $xF_3$ data only, whereas no nuclear corrections are applied to the
$F_2^p$, $F_2^d$ datasets.
\section{Fit Results}\label{sec4}

The global QCD analysis presented in this work is based on a simultaneous fit to the experimental data for the structure functions $xF_3$, $F_2^p$, $F_2^d$, and $F_2^{\rm NS}$. The datasets, their kinematic coverage, and the numbers of data points before and after the applied cuts are summarized in Table~\ref{tab:combined_data_cuts}. The $xF_3$ data are taken from the CCFR, NuTeV, and CHORUS experiments and comprise 230 points, all of which are retained after the cuts. The proton structure-function data, $F_2^p$, are drawn from BCDMS, NMC, SLAC, H1, and ZEUS and consist of 467 points before cuts, with 248 points included in the fit. The deuteron data, $F_2^d$, come from BCDMS, NMC, and SLAC, yielding 232 points before cuts and 159 after cuts. The non-singlet dataset, $F_2^{\rm NS}$, also includes measurements from BCDMS, NMC, and SLAC, with 208 points before cuts and 165 retained in the analysis. Nuclear correction factors are applied to the $xF_3$ neutrino--nucleus data only; no nuclear corrections are applied to the $F_2^p$, $F_2^d$, or $F_2^{\rm NS}$ datasets. All four datasets are included simultaneously in the global fit to constrain the valence-quark distributions.

To quantify the agreement between the theoretical predictions and the experimental measurements, we use the standard $\chi^2$ definition
\begin{equation}
	\chi_{\mathrm{global}}^2
	=
	\sum_{n}\chi_n^2
	=
	\sum_{n}\sum_{i}
	\left(
	\frac{F_{i,n}^{\mathrm{data}}-F_{i,n}^{\mathrm{theor}}}
	{\Delta F_{i,n}^{\mathrm{data}}}
	\right)^2 ,
	\label{eq:Chi2global}
\end{equation}
where $n$ labels the experimental datasets included in the global fit and $i$ labels the data points retained after the applied cuts. Here, $F_{i,n}^{\mathrm{data}}$ and $F_{i,n}^{\mathrm{theor}}$ denote the measured value and the corresponding theoretical prediction, respectively, while $\Delta F_{i,n}^{\mathrm{data}}$ is the total experimental uncertainty, obtained by adding the statistical and systematic uncertainties in quadrature.To estimate the error bands and parameter uncertainties, we have employed the Hessian method (see, e.g., Ref.~\cite{Stump:2001gu}).
The structure function $F$ denotes the relevant observable in each dataset: $xF_3$, $F_2^p$, $F_2^d$, or $F_2^{\rm NS}$. The sum runs over all datasets included in the global analysis, namely those from CCFR, NuTeV, CHORUS, BCDMS, NMC, SLAC, H1, and ZEUS. The free parameters of the valence-quark distributions are determined by minimizing $\chi_{\mathrm{global}}^2$ using the {\tt MINUIT} package~\cite{MINUIT}.

\begin{table}[htbp]
	\centering
	\caption{Experimental data points and cuts for $xF_3$, $F_2^p$, $F_2^d$,
		and $F_2^{\rm NS}$.}
	\label{tab:combined_data_cuts}
	\small
	\resizebox{\textwidth}{!}{%
		\begin{tabular}{|l|l|c|c|c|c|}
			\hline
			\textbf{Structure function} &
			\textbf{Experiment} &
			\textbf{$x$ range} &
			\textbf{$Q^2$ range ($\mathrm{GeV}^2$)} &
			\textbf{Data points} &
			\textbf{Data points after cuts} \\
			\hline
			
			$xF_3$ & CCFR
			& $0.0075 \leq x \leq 0.75$
			& $1.3 \leq Q^2 \leq 125.9$
			& 116 & 116 \\
			
			$xF_3$ & NuTeV
			& $0.015 \leq x \leq 0.75$
			& $3.162 \leq Q^2 \leq 50.118$
			& 64 & 64 \\
			
			$xF_3$ & CHORUS
			& $0.02 \leq x \leq 0.65$
			& $2.052 \leq Q^2 \leq 81.55$
			& 50 & 50 \\
			
				$xF_3$ & Nuclear (total)
			& -- & --
			& 230 & 230 \\
			\hline
			\hline
			
			$F_2^p$ & BCDMS (100)
			& $0.35\leq x \leq0.75$ & $11.75\leq Q^2 \leq75.00$
			& 51 & 29 \\
			
			$F_2^p$ & BCDMS (120)
			& $0.35\leq x \leq0.75$ & $13.25\leq Q^2 \leq75.00$
			& 59 & 32 \\
			
			$F_2^p$ & BCDMS (200)
			& $0.35\leq x \leq0.75$ & $32.50\leq Q^2 \leq137.50$
			& 50 & 28 \\
			
			$F_2^p$ & BCDMS (280)
			& $0.35\leq x \leq0.75$ & $43.00\leq Q^2 \leq230.00$
			& 49 & 26 \\
			
			$F_2^p$ & NMC (comb)
			& $0.35\leq x \leq0.50$ & $7.00\leq Q^2 \leq65.00$
			& 15 & 14 \\
			
			$F_2^p$ & SLAC (comb)
			& $0.30\leq x \leq0.62$ & $7.30\leq Q^2 \leq21.39$
			& 57 & 57 \\
			
			$F_2^p$ & H1 (hQ2)
			& $0.40\leq x \leq0.65$ & $200\leq Q^2 \leq30000$
			& 26 & 26 \\
			
			$F_2^p$ & ZEUS (hQ2)
			& $0.40\leq x \leq0.65$ & $650\leq Q^2 \leq30000$
			& 15 & 15 \\

			$F_2^p$ & Proton (total)
			& -- & --
			& 322 & 227 \\
			\hline
			
			$F_2^d$ & BCDMS (120)
			& $0.35\leq x \leq0.75$ & $13.25\leq Q^2 \leq99.00$
			& 59 & 32 \\
			
			$F_2^d$ & BCDMS (200)
			& $0.35\leq x \leq0.75$ & $32.50\leq Q^2 \leq137.50$
			& 50 & 28 \\
			
			$F_2^d$ & BCDMS (280)
			& $0.35\leq x \leq0.75$ & $43.00\leq Q^2 \leq230.00$
			& 49 & 26 \\
			
			$F_2^d$ & NMC (comb)
			& $0.35\leq x \leq0.50$ & $7.00\leq Q^2 \leq65.00$
			& 15 & 14 \\
			
			$F_2^d$ & SLAC (comb)
			& $0.30\leq x \leq0.62$ & $10.00\leq Q^2 \leq21.40$
			& 59 & 59 \\
			
			$F_2^d$ & Deuteron (total)
			& -- & --
			& 232 & 159 \\
			\hline
			
			$F_2^{\rm NS}$ & BCDMS (120)
			& $0.070\leq x \leq0.275$ & $8.75\leq Q^2 \leq43.00$
			& 36 & 30 \\
			
			$F_2^{\rm NS}$ & BCDMS (200)
			& $0.070\leq x \leq0.275$ & $17.00\leq Q^2 \leq75.00$
			& 29 & 28 \\
			
			$F_2^{\rm NS}$ & BCDMS (280)
			& $0.100\leq x \leq0.275$ & $32.50\leq Q^2 \leq115.50$
			& 27 & 26 \\
			
			$F_2^{\rm NS}$ & NMC (comb)
			& $0.013\leq x \leq0.275$ & $4.50\leq Q^2 \leq65.00$
			& 88 & 53 \\
			
			$F_2^{\rm NS}$ & SLAC (comb)
			& $0.153\leq x \leq0.293$ & $4.18\leq Q^2 \leq5.50$
			& 28 & 28 \\
			
			$F_2^{\rm NS}$ & Nonsinglet (total)
			& -- & --
			& 208 & 165 \\
			\hline
		\end{tabular}%
	}
\end{table}

\begin{table*}[t]
	\centering
	\caption{Best-fit parameters and corresponding uncertainties at the initial scale $Q_0^2 = 1.0$~GeV$^2$ obtained from the Laplace transform analysis at NLO and NNLO (N$^2$LO).}
	\label{ta2}
	\begin{tabular}{ccc}
		\hline
		Parameter & NLO & NNLO (N$^2$LO) \\ \hline
		$a_{u}^{1}$              & $-0.29841 \pm 0.000243704$  & $-0.25709 \pm 0.00105682$  \\ 
		$a_{u}^{2}$              & $0.055435 \pm 0000106118$   & $0.040104 \pm 0.0044080$ \\ 
		$a_{u}^{3}$              & $-0.0052468 \pm 0.0000268904$  & $-0.00303544 \pm0.000088380$  \\ 
		$\alpha_{u_{v}}$         & $0.289981 \pm  0.000208101$   & $0.38717 \pm0.0012371$  \\ 
		$\beta_{u_{v}}$          & $5.0000034 \pm 0.000223021$   & $4.3389 \pm 0.018919$  \\ 
		$\beta_{d_{v}}$          & $0.39972 \pm 0.00105178$  & $1.28106 \pm 0.0419647$  \\ 
		$\Lambda$  & $0.29982 \pm 0.0059273$   & $ 0.400972 \pm  0.00997265$  \\ 
		$\alpha_{s}(M_{z}^{2})$  & $0.117548 \pm 0.000344$   & $0.113172 \pm 0.0004393$  \\ \hline
		$\chi^{2}/\text{d.o.f}$  & $ 5634.985 / 774 =7.28034$   & $ 4834.605 / 774 =6.2462$  \\ \hline
	\end{tabular}
\end{table*}

Table~\ref{ta2} summarizes the best-fit parameter values
and their uncertainties obtained at NLO and NNLO. In addition to the standard kinematic cuts ($Q^2 \geq 4\,\mathrm{GeV}^2$ and $W^2 \geq 12.5\,\mathrm{GeV}^2$) applied in our analysis, we also take into account target mass corrections (TMC) for the proton $F_2^p(x, Q^2)$ and deuteron $F_2^d(x, Q^2)$ structure functions in the kinematic region where $Q^2 \geq 4\,\mathrm{GeV}^2$ and $4 < W^2 < 12.5\,\mathrm{GeV}^2$. For this purpose, we extend our QCD fit results to this region by incorporating the target mass correction framework.
 The fitted parameters
include the three Gegenbauer coefficients $a_u^1$, $a_u^2$, and $a_u^3$;
the exponents $\alpha_{u_v}$, $\beta_{u_v}$, and $\beta_{d_v}$ that
control the asymptotic behavior of the valence-quark distributions; and
the QCD scale parameter $\Lambda$. The corresponding values of
$\alpha_s(M_Z^2)$ and the goodness-of-fit statistic are also reported.

The NLO and NNLO results show moderate shifts in several fitted
parameters. In particular, the central values of the Gegenbauer
coefficients change between the two perturbative orders, while
$\alpha_{u_v}$ and $\beta_{u_v}$ remain positive at both orders.
The fitted value of $\beta_{d_v}$ is also positive at NLO and NNLO,
with a larger central value at NNLO. Thus, within the parameterization
used here, the large-$x$ behavior of the down-valence distribution is
controlled by a positive exponent. The fitted QCD scale parameter
increases from $\Lambda=0.29982\pm0.0059273$ at NLO to
$\Lambda=0.400972\pm0.00997265$ at NNLO.

The extracted strong-coupling values are
$\alpha_s(M_Z^2)=0.117548\pm0.000344$ at NLO and
$\alpha_s(M_Z^2)=0.113172\pm0.0004393$ at NNLO. Both are compatible,
within the stated uncertainties and perturbative-order dependence, with
the world-average value
$\alpha_s(M_Z^2)=0.1161\pm0.00149$~\cite{ParticleDataGroup:2014cgo}. The fitted
goodness-of-fit values are
$\chi^2/\mathrm{d.o.f.}=5634.985/774=7.28034$ at NLO and
$\chi^2/\mathrm{d.o.f.}=4834.605/774=6.2462$ at NNLO. The lower
NNLO value indicates an improved description of the fitted data relative
to NLO, although both values remain substantially above unity. This
suggests that the present theoretical description and/or the treatment
of the experimental uncertainties does not fully account for the
observed data, and that further investigation is warranted.

\begin{figure}[htbp]
	\centering
	
	\includegraphics[width=0.48\textwidth]{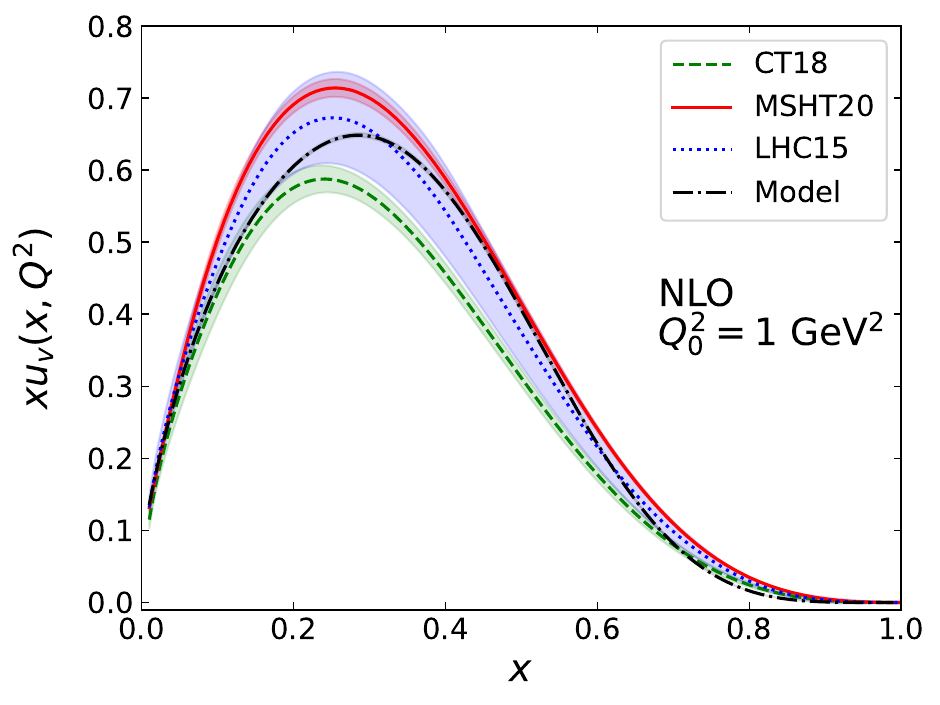}\hfill
	\includegraphics[width=0.48\textwidth]{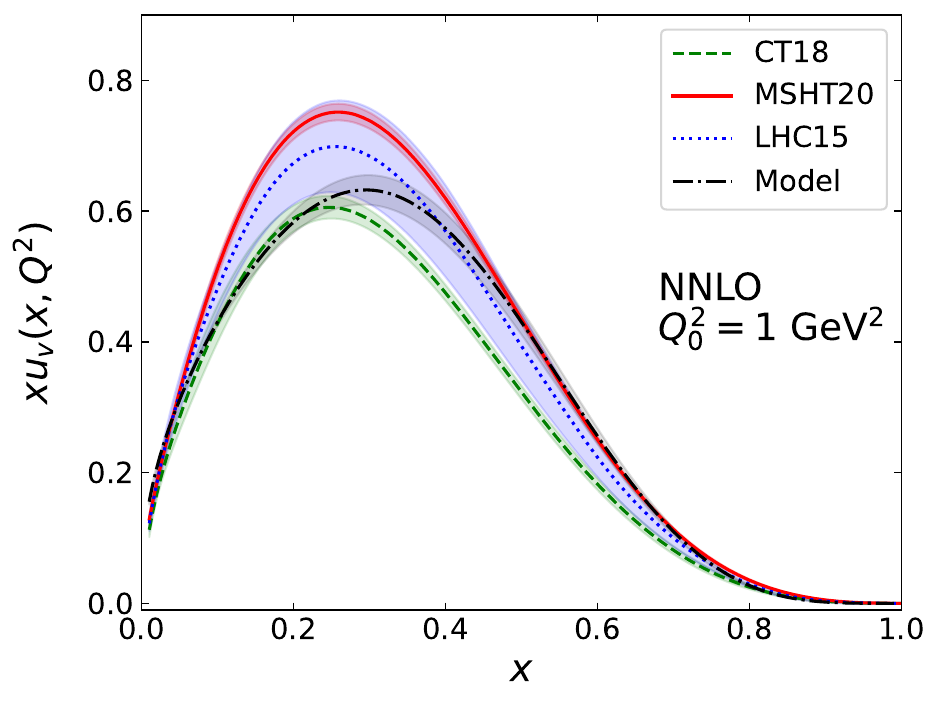}
	
	\medskip
	
	\includegraphics[width=0.48\textwidth]{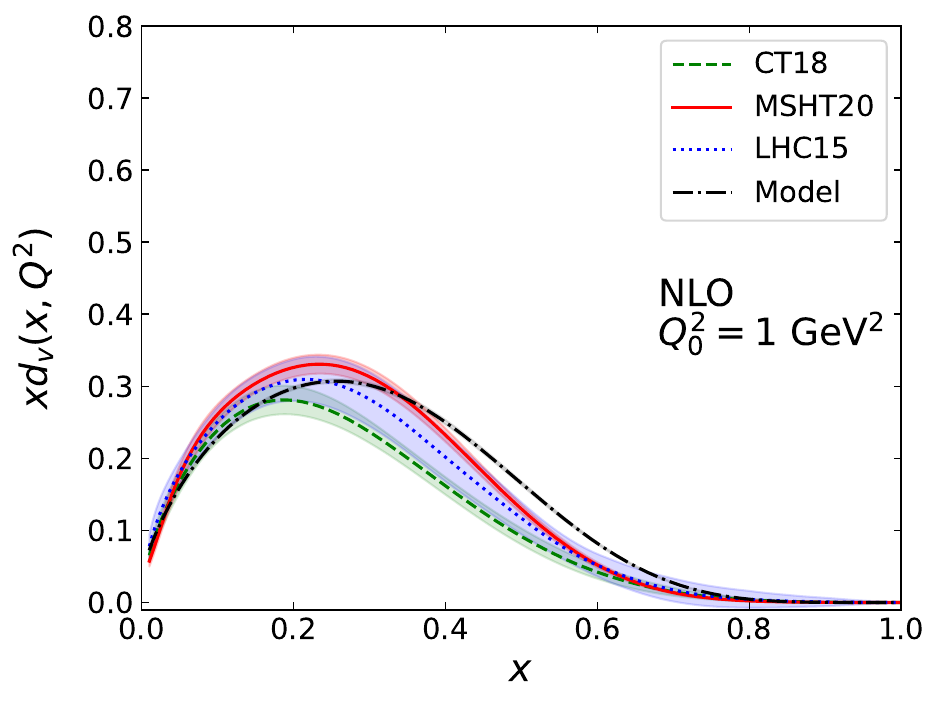}\hfill
	\includegraphics[width=0.48\textwidth]{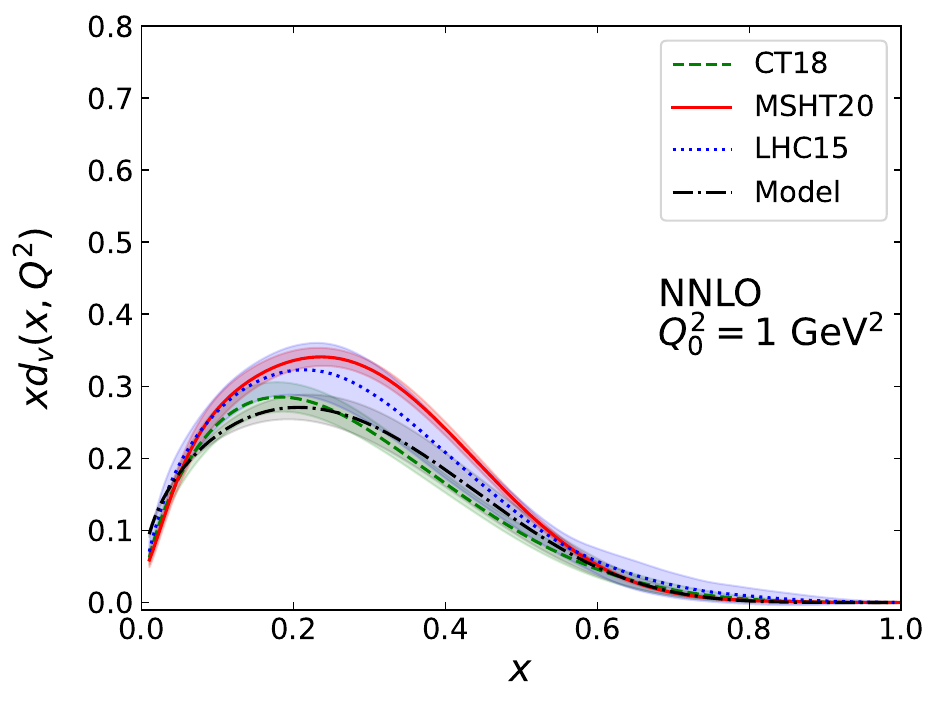}
	
	\caption{\sf The $xu_v$ and $xd_v$ at $Q_0^2=1\,\mathrm{GeV}^2$ in the Laplace analysis for the NLO and NNLO models compare with CT18~\cite{Hou:2019efy} , MSHT20~\cite{Cridge:2021qfd} and LHC15~\cite{Butterworth:2015oua}.}
	\label{fig1-1}
\end{figure}

Figure~\ref{fig1-1} shows the up- and down-valence distributions,
$x u_v(x,Q^2)$ and $x d_v(x,Q^2)$, at the initial scale
$Q_0^2=1\,\mathrm{GeV}^2$, obtained from the present Laplace-space
analysis with a Gegenbauer-polynomial parametrization. The results are
shown at NLO and NNLO and compared with the CT18~\cite{Hou:2019efy},
MSHT20~\cite{Cridge:2021qfd}, and LHC15~\cite{Butterworth:2015oua}
PDF sets. The shaded regions indicate the uncertainty bands of the
corresponding distributions.

For $x u_v$, the present NLO and NNLO results reproduce the characteristic
valence-quark shape: the distribution rises at small $x$, reaches a
maximum at intermediate $x$, and decreases towards large $x$. The
predicted curves and their uncertainty bands show a degree of overlap
with the reference PDF sets, although differences in their central
values are visible, particularly around the peak region. The comparison
also shows that the relative ordering of the predictions depends on the
perturbative order and on $x$.

The $x d_v$ distributions exhibit a similar rise and fall, but their
peak values are smaller than those of $x u_v$. The spread among the
different predictions is more noticeable over parts of the intermediate-
and large-$x$ regions. This behavior reflects the different constraints
and parametrizations entering the global PDF analyses, as well as the
sensitivity of the down-valence distribution to the fitted data.

Overall, the figure provides a comparison of the valence distributions
extracted in the present analysis with those of established PDF sets.
The overlap of the uncertainty bands in several $x$ regions indicates
where the results are compatible within their quoted uncertainties,
while the remaining differences highlight the dependence on the
perturbative order and the details of the PDF parametrization.

\begin{figure}[htbp]
	\centering
	
	\includegraphics[width=0.48\textwidth]{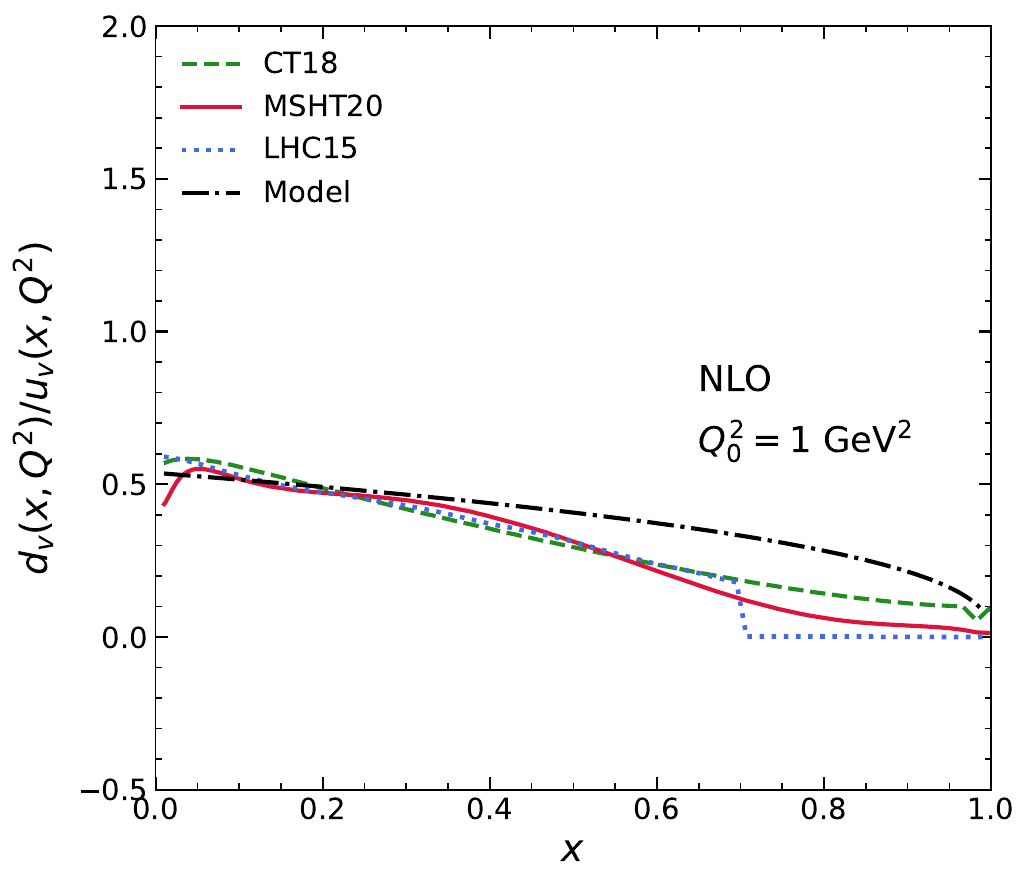}\hfill
	\includegraphics[width=0.48\textwidth]{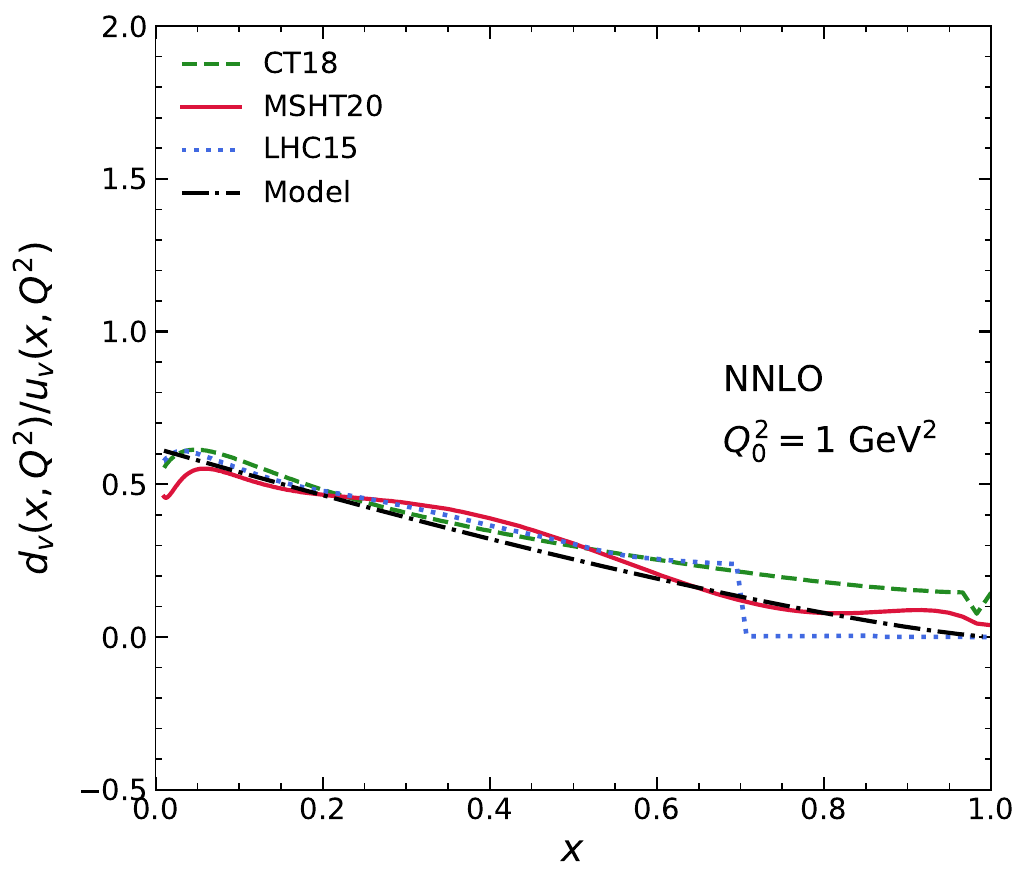}

	\caption{\sf The ratio $d_v/u_v$  at $Q_0^2=1\,\mathrm{GeV}^2$ in the Laplace analysis for the NLO and NNLO models compare with CT18~\cite{Hou:2019efy} , MSHT20~\cite{Cridge:2021qfd} and LHC15~\cite{Butterworth:2015oua}.}
	\label{fig1-2}
\end{figure}

Figure~\ref{fig1-2} shows the ratio of the down- to up-valence
distributions, $d_v(x,Q^2)/u_v(x,Q^2)$, at the initial scale
$Q_0^2=1\,\mathrm{GeV}^2$. The predictions obtained in the present
Laplace-space analysis are shown at NLO and NNLO and compared with the
CT18~\cite{Hou:2019efy}, MSHT20~\cite{Cridge:2021qfd}, and
LHC15~\cite{Butterworth:2015oua} parametrizations.

At small and intermediate $x$, the model predictions show broadly
similar behavior to the reference parametrizations. The ratio varies
moderately across this region, although differences among the central
curves are visible. At larger $x$, the predictions separate more
noticeably: the model ratio remains above some of the reference curves
over part of the range, while the different parametrizations exhibit
distinct endpoint behavior as $x$ approaches unity. The comparison
between NLO and NNLO also shows that the ratio depends on the
perturbative order, particularly at large $x$.

These differences at large $x$ should be interpreted with care, since
the ratio there is sensitive to the endpoint behavior of the individual
valence distributions and to the flexibility of their parametrizations.
Thus, while the figure allows a direct comparison of the
$d_v/u_v$ ratio across perturbative orders and PDF sets, the visible
differences at high $x$ do not by themselves establish a disagreement
in the data-constrained region.

Figure~\ref{fig1} displays the extracted valence quark distributions $x u_v$ and $x d_v$ at the initial scale $Q_0^2 = 1~\text{GeV}^2$ for both NLO and NNLO. The distributions exhibit the expected qualitative behavior: the up-valence distribution dominates the down-valence at all $x$ values, reflecting the proton's valence structure with two up quarks and one down quark. The peak positions and widths of the distributions are consistent with typical PDF parameterizations, with the up-valence distribution peaking at slightly lower $x$ compared to the down-valence distribution. The Gegenbauer expansion provides sufficient shape flexibility, particularly in the intermediate-$x$ region where the data impose stringent constraints, while the asymptotic exponents control the behavior at the kinematic endpoints. The differences between the NLO and NNLO predictions are minimal, indicating that the perturbative convergence is well under control and that the extracted valence distributions are robust against higher-order QCD corrections.

\begin{figure}[H]
	\begin{center}
	\includegraphics[clip,width=0.7\textwidth]{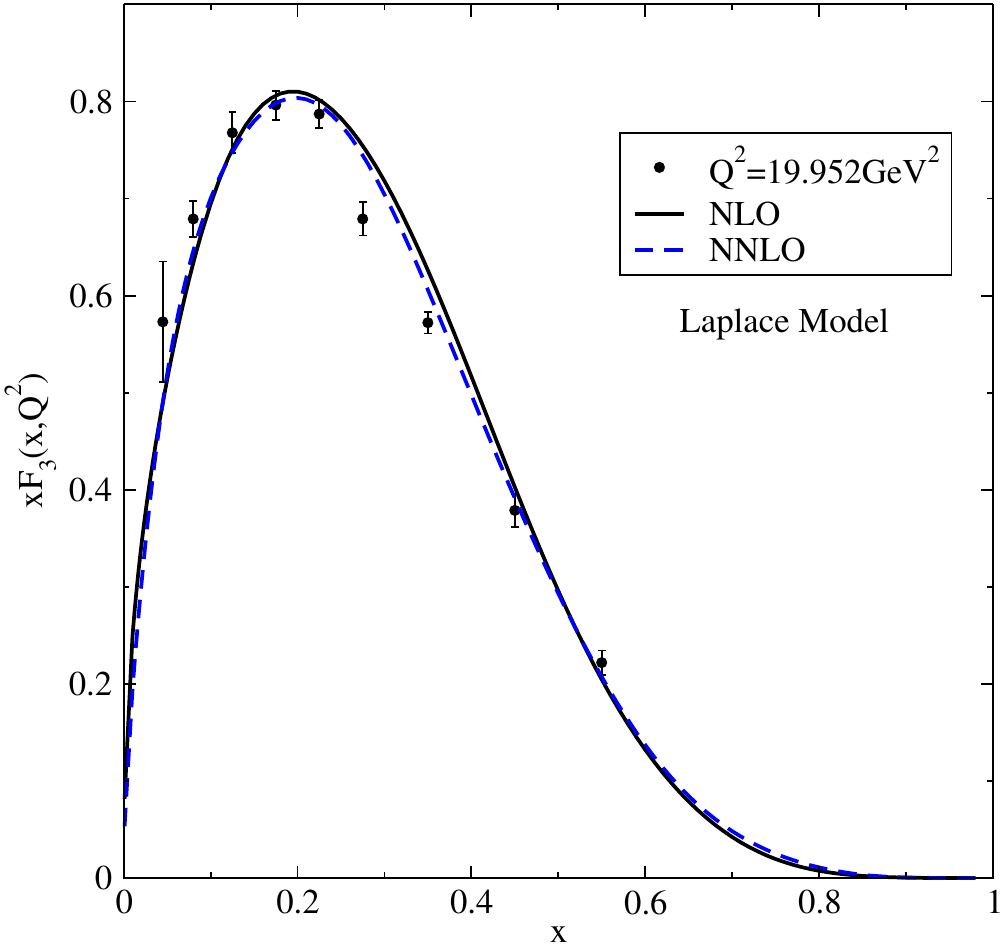}%
	\caption{\sf The $xF_3$  $Q^2=19.952GeV^2$ in Laplace analysis in NLO and NNLO model } \label{fig2}
	\end{center}
\end{figure}
\begin{figure}[htbp]
	\centering
	\includegraphics[clip,width=0.32\textwidth]{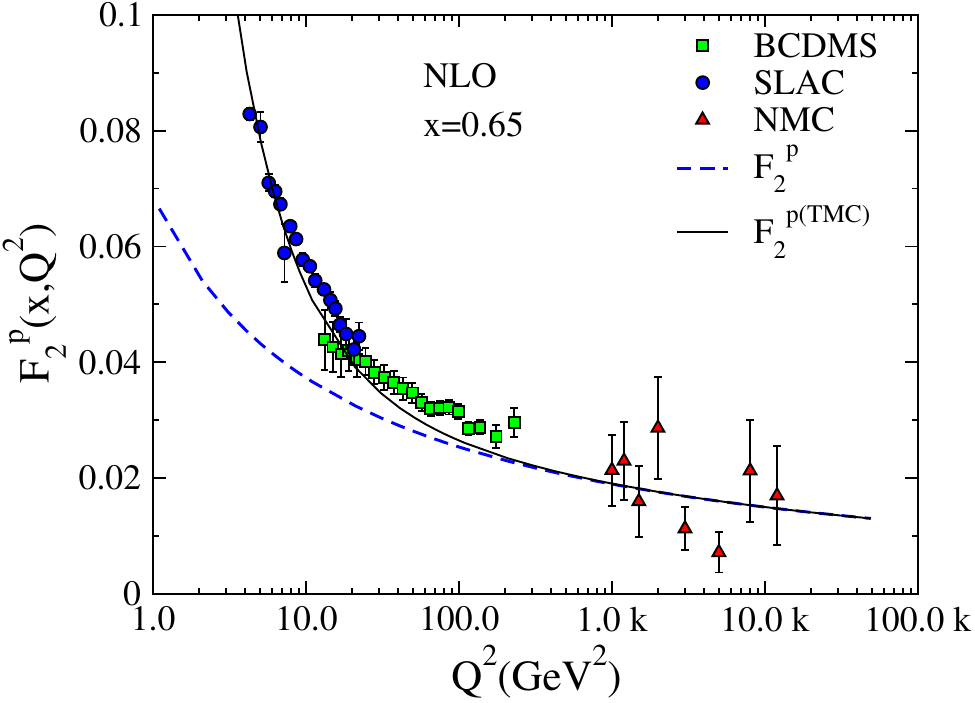}\hfill
	\includegraphics[clip,width=0.32\textwidth]{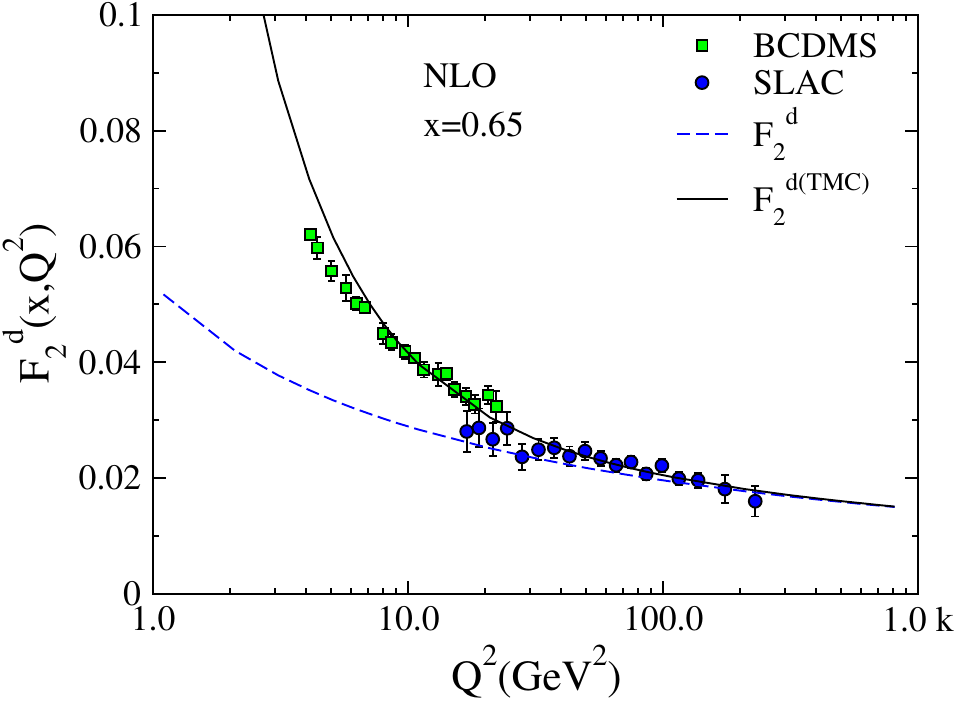}\hfill
	\includegraphics[clip,width=0.32\textwidth]{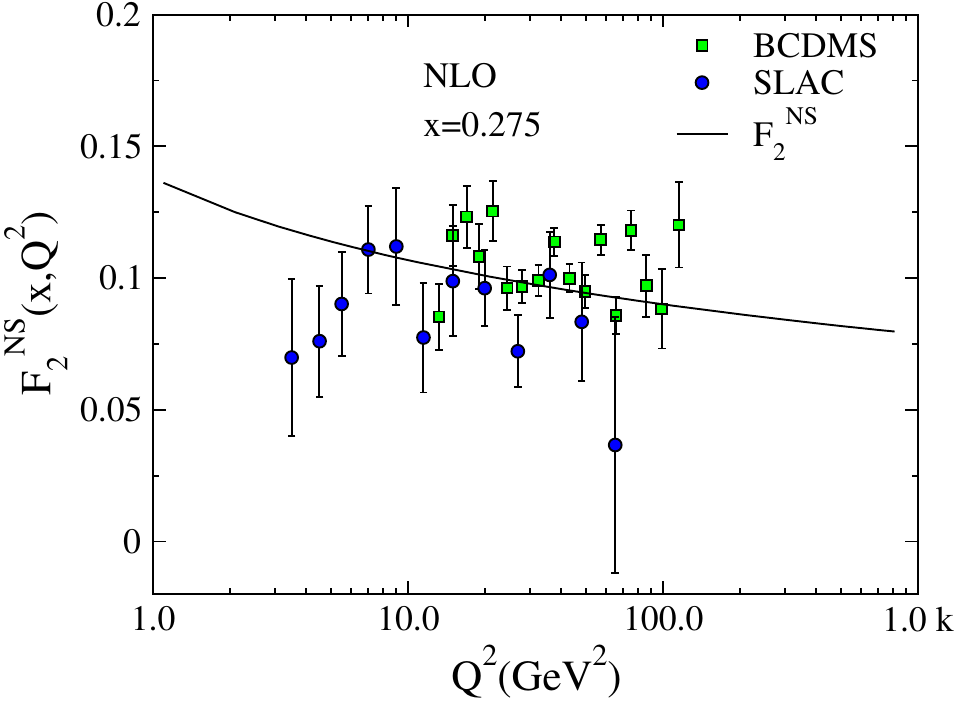}\\[2mm]
	\includegraphics[clip,width=0.32\textwidth]{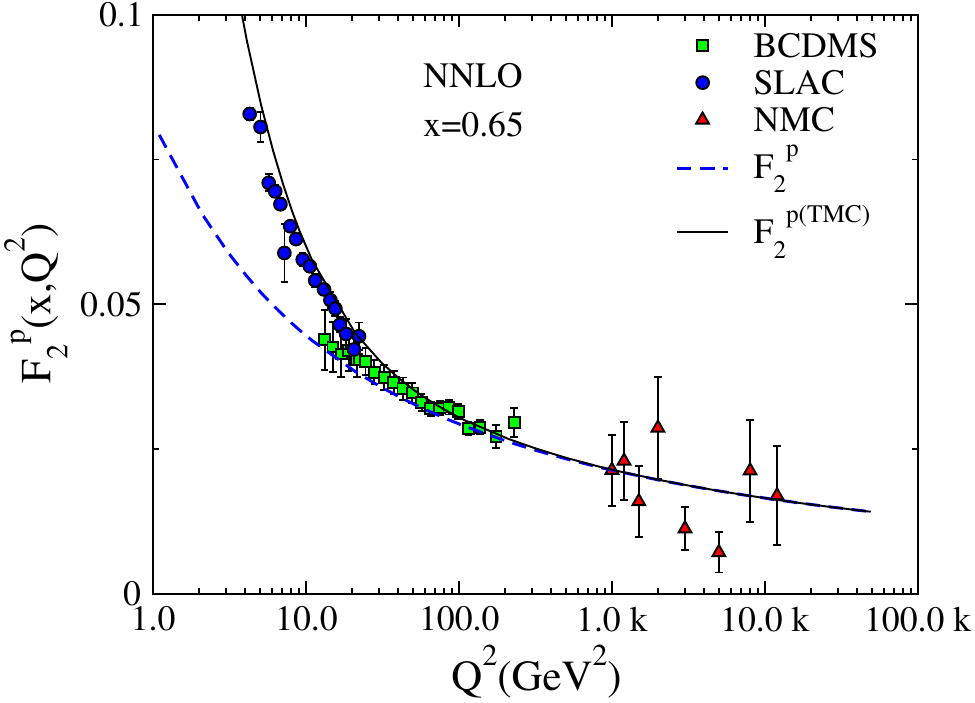}\hfill
	\includegraphics[clip,width=0.32\textwidth]{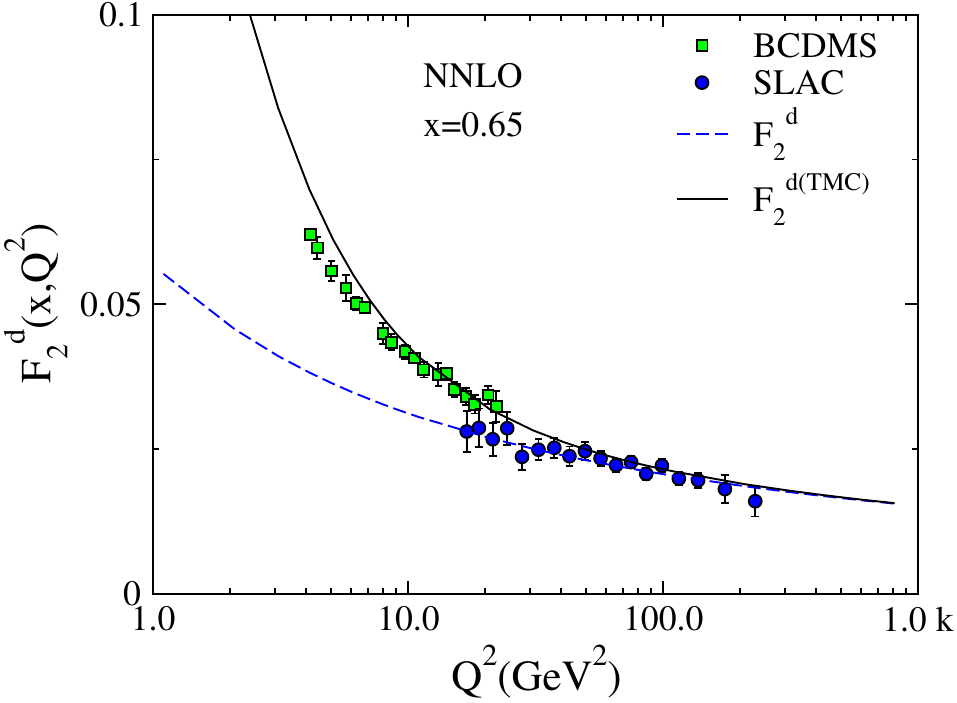}\hfill
	\includegraphics[clip,width=0.32\textwidth]{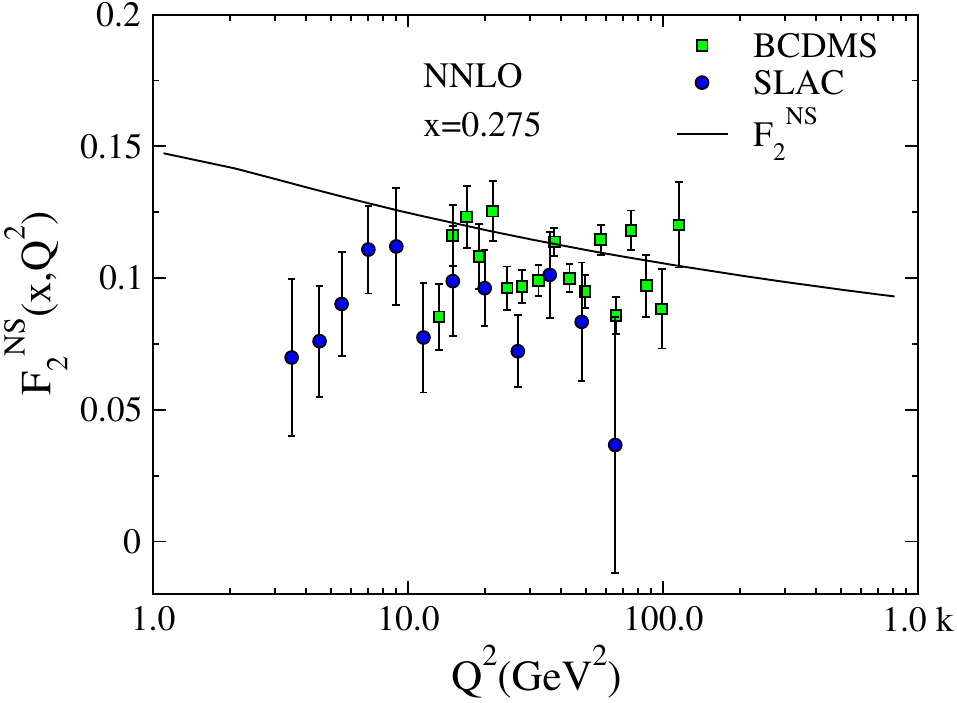}
	\caption{The structure functions $F_2^p$, $F_2^d$, and $F_2^{\rm NS}$ as functions of $Q^2$ (in $\mathrm{GeV}^2$), shown in the left, center, and right columns, respectively. The corresponding Bjorken-$x$ values are $0.65$, $0.55$, and $0.275$. The upper and lower rows show the NLO and NNLO results, respectively. Experimental data are from BCDMS (green squares), SLAC (blue circles), and NMC (red triangles, where shown). Solid black curves represent the total theoretical predictions, while blue dashed curves indicate the target-mass-correction (TMC) contributions where displayed. The horizontal axes are logarithmic in $Q^2$.}
	
	\label{fig22}
\end{figure}

Figure~\ref{fig2} illustrates the comparison between the theoretical predictions and the experimental data for the structure function $xF_3$ at a representative scale of $Q^2 = 19.952~\text{GeV}^2$. At this momentum transfer, the perturbative QCD evolution has significantly altered the shape of the structure function relative to its input form, particularly in the small-$x$ region where the valence distributions are depleted through DGLAP evolution. The theoretical curves are computed using the best-fit parameters from Table~\ref{ta2}, and they demonstrate a satisfactory agreement with the experimental data across the entire measured $x$ range. The NLO and NNLO curves are almost indistinguishable at this scale, indicating that the perturbative series for the non-singlet structure function is well behaved and that the dominant uncertainty stems from the input parameterization rather than from the truncation of the perturbative expansion. The visible scatter in the high-$x$ region reflects the aforementioned difficulty in describing the NuTeV and CHORUS data simultaneously, a feature that will be further investigated through the sum rule analysis presented in the subsequent sections. Overall, the Laplace-transform framework combined with the Gegenbauer parameterization and dynamic nuclear corrections yields a flexible and efficient description of the $xF_3$ data, demonstrating its viability for precision QCD phenomenology.

Figure~\ref{fig22} compares the measured
$Q^2$-dependence of the proton, deuteron, and nonsinglet structure
functions with the predictions of the present analysis. The proton
structure function $F_2^p$ is shown in the left column at
$x=0.65$, the deuteron structure function $F_2^d$ in the center
column at $x=0.55$, and the nonsinglet structure function
$F_2^{\rm NS}$ in the right column at $x=0.275$. The upper and lower
rows correspond to the NLO and NNLO results, respectively. The
experimental data are from BCDMS, SLAC, and, where shown, NMC.

For $F_2^p$ and $F_2^d$, the solid curves show the theoretical
predictions including target-mass corrections (TMCs), while the blue
dashed curves show the corresponding results without TMCs. The
comparison illustrates the importance of target-mass effects in the
large-$x$ region, particularly at lower and moderate values of $Q^2$.
As $Q^2$ increases, the difference between the predictions with and
without TMCs becomes smaller, as expected for power-suppressed
corrections. The NLO and NNLO curves describe the overall
$Q^2$-dependence of the proton and deuteron data, with the NNLO
predictions providing the corresponding higher-order results.

The right column shows the $F_2^{\rm NS}$ data and theoretical
predictions at $x=0.275$. In this case, the solid curves represent the
calculated structure function; no separate TMC curve is displayed.
The data exhibit noticeable scatter around the predicted
$Q^2$-dependence, which is also reflected in the difference between
the NLO and NNLO results.

Taken together, these panels compare the perturbative-order dependence
of the calculation with the measured structure functions used in the
global analysis. The figure also demonstrates that the treatment of
target-mass corrections is relevant to the description of the
large-$x$ proton and deuteron data. The horizontal axes are logarithmic
in $Q^2$.

\begin{figure}[H]
	\centering
	\includegraphics[clip,width=0.82\textwidth]{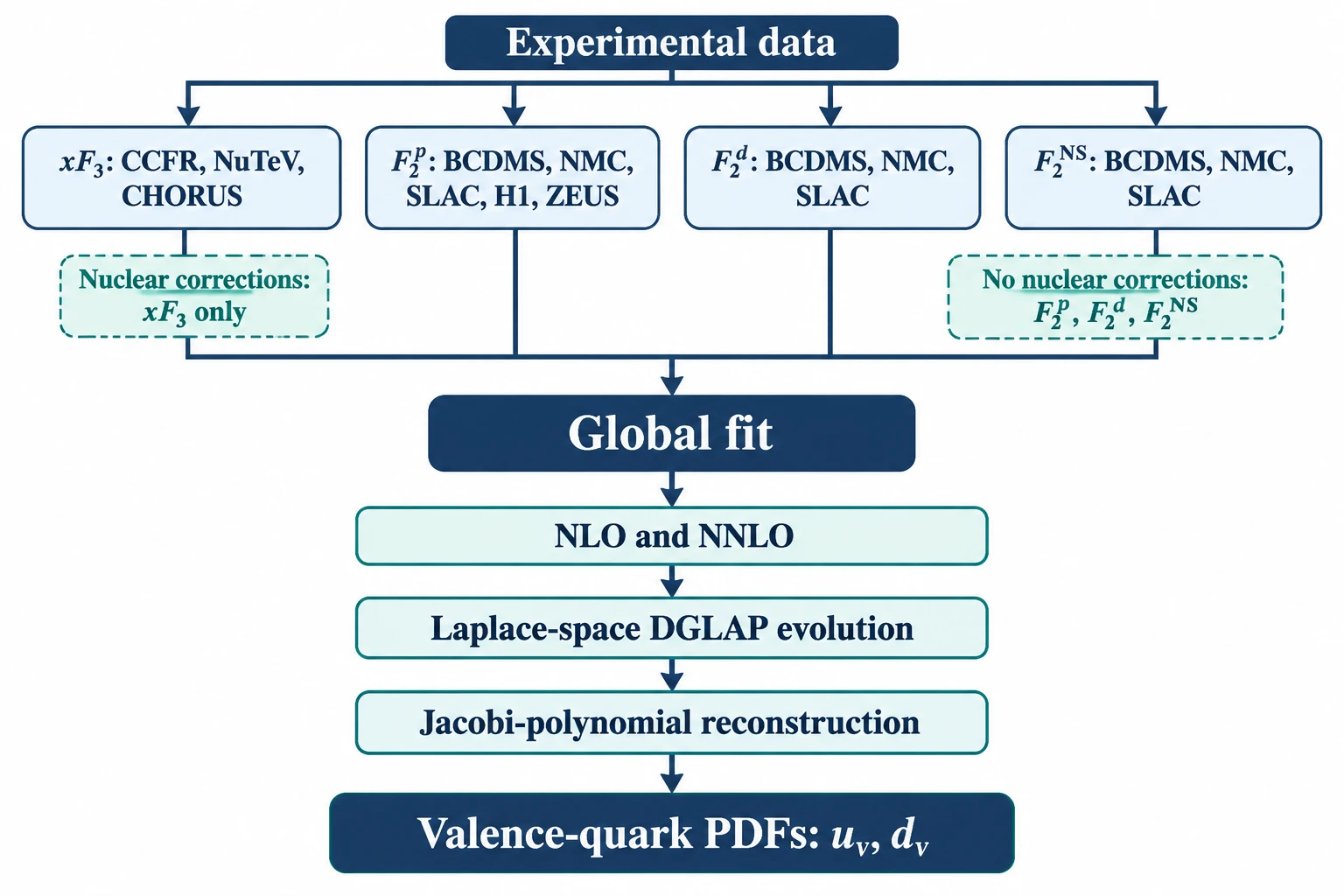}\hfill
\caption{Schematic overview of the global analysis. The experimental data include
	$xF_3$ from CCFR, NuTeV, and CHORUS; $F_2^p$ from BCDMS, NMC, SLAC, H1, and ZEUS;
	and $F_2^d$ and $F_2^{\mathrm{NS}}$ from BCDMS, NMC, and SLAC. Nuclear corrections
	are included for the $xF_3$ data only, while no nuclear corrections are applied
	to the $F_2^p$, $F_2^d$, or $F_2^{\mathrm{NS}}$ data. All data sets are analyzed
	simultaneously in a global fit at NLO and NNLO. The parton distributions are
	evolved using the DGLAP equations in Laplace space, and the resulting structure
	functions are reconstructed using the Jacobi-polynomial method. The analysis
	constrains the valence-quark distributions $u_v$ and $d_v$.}
\label{fig:analysis_flowchart}
	
	\label{fig22}
\end{figure}

Figure~\ref{fig:analysis_flowchart} summarizes the main steps of the
analysis. Four sets of deep-inelastic scattering structure-function data
are included in a simultaneous global fit: $xF_3$, $F_2^p$, $F_2^d$, and
$F_2^{\mathrm{NS}}$. Nuclear corrections are applied to the $xF_3$ data,
whereas the proton, deuteron, and nonsinglet $F_2$ data are included
without nuclear corrections. The fitted distributions are analyzed at
both NLO and NNLO. Their scale dependence is calculated by solving the
DGLAP evolution equations in Laplace space, and the structure functions
are reconstructed in Bjorken-$x$ space using the Jacobi-polynomial
expansion. This procedure provides the input for the simultaneous
determination of the valence-quark distributions $u_v$ and $d_v$.

\section{Sum Rules: Theoretical Framework and Numerical Results}\label{sec5}

The investigation of fundamental sum rules constitutes an essential component of any QCD analysis based on deep-inelastic scattering data. These sum rules, derived from current algebra, isospin symmetry, and the operator product expansion, provide stringent tests of the internal consistency of parton distribution functions and serve as sensitive probes of both perturbative and non-perturbative QCD dynamics. In this work, we examine three benchmark sum rules within the Laplace-transform framework: the Gross–Llewellyn Smith (GLS), Bjorken, and Adler sum rules. A distinctive feature of our analysis is the systematic incorporation of nuclear medium effects, which are essential for interpreting data obtained from heavy targets such as iron and lead. For completeness and to quantify the impact of nuclear corrections, we also evaluate these sum rules in the idealized free-proton limit, thereby isolating the genuine nuclear modifications from the baseline QCD predictions.

The Adler and GLS sum rules are particularly relevant for charged-current neutrino scattering, as they involve the parity-violating structure function $F_3$, which is purely non-singlet in nature and thus directly sensitive to valence quark distributions. The Bjorken sum rule, although originally formulated for polarized electron scattering, possesses an unpolarized analog in neutrino scattering that shares similar non-singlet characteristics. By comparing the nuclear and free-nucleon results, we can assess the magnitude of nuclear effects and their influence on the extracted values of fundamental constants such as $\alpha_s(M_Z^2)$ and the nucleon axial charge $g_A$. Below, we present the theoretical formulation and numerical results for each sum rule in turn.

\subsection{Adler Sum Rule}

The Adler sum rule represents a unique and exact relation in charged-current neutrino–nucleon deep-inelastic scattering, derived rigorously from current algebra and isospin symmetry without recourse to perturbative QCD approximations. Unlike the GLS or Bjorken sum rules, which receive perturbative corrections, the Adler sum rule is \textit{exact} to all orders in $\alpha_s$, making it a pristine probe of valence quark normalization and nuclear isoscalarity.

In the naive parton model, this sum rule relates the difference between neutrino and antineutrino structure functions directly to the net valence quark content of the target. For a free proton, the sum rule is conventionally expressed as~\cite{Adler:2009dw}:

\begin{equation}
	\int_{0}^{1} \frac{dx}{x} \left[ F_2^{\bar{\nu}p}(x, Q^2) - F_2^{\nu p}(x, Q^2) \right] = 2,
\end{equation}

which, in terms of valence quark distributions, becomes:

\begin{equation}
	\int_{0}^{1} \left[ u_v^p(x) - d_v^p(x) \right] dx 
	= \int_{0}^{1} \left[ (u(x) - \bar{u}(x)) - (d(x) - \bar{d}(x)) \right] dx = 1.
\end{equation}

The factor of 2 in Eq.~(1) originates from the overall charged-current normalization, yielding the well-known result $S_{\rm Ad}^{p} = 2$ for the proton target.

For a nuclear target characterized by atomic number $Z$, neutron number $N$, and mass number $A = Z + N$, the total valence content is obtained by integrating the nuclear PDFs:

\begin{equation}
	\int_0^1 u_v^A(x) \, dx = 2Z + N, \qquad 
	\int_0^1 d_v^A(x) \, dx = Z + 2N.
\end{equation}

Subtracting these two relations yields the net valence difference for the entire nucleus:

\begin{equation}
	\int_0^1 \left[ u_v^A(x) - d_v^A(x) \right] dx = Z - N.
\end{equation}

Consequently, the Adler sum rule for the full nucleus $A$ takes the form:

\begin{equation}
	\int_{0}^{1} \frac{dx}{x} \left[ F_2^{\bar{\nu}A}(x, Q^2) - F_2^{\nu A}(x, Q^2) \right] = 2(Z - N).
\end{equation}

To facilitate comparison with free-nucleon results, we normalize the nuclear structure functions by the mass number $A$, leading to the per-nucleon Adler sum rule \cite{Kulagin:2007ju}:

\begin{equation}
	\frac{1}{A} \int_{0}^{1} \frac{dx}{x} \left[ F_2^{\bar{\nu}A}(x, Q^2) - F_2^{\nu A}(x, Q^2) \right] = 2 \left( \frac{Z - N}{A} \right).
\end{equation}

Equation~(6) reveals that the per-nucleon Adler sum rule is directly proportional to the non-isoscalarity factor, defined as:

\begin{equation}\label{eq:ad}
	\delta_{\text{ns}} \equiv \frac{Z - N}{A}.
\end{equation}

For an isoscalar nucleus ($Z = N$), the right-hand side of Eq.~(6) vanishes identically. However, for non-isoscalar nuclei such as iron ($^{56}\mathrm{Fe}$ with $Z=26$, $N=30$), the non-isoscalarity factor is non-zero:

\begin{equation}
	\delta_{\text{ns}} \approx -0.0714,
\end{equation}

At the reference scale $Q^2=3\,\mathrm{GeV}^2$, the exact nuclear
Adler sum rule for the idealized proton--neutron composition considered
here is
\begin{equation}
	S_{\mathrm{Ad}}^{A,\mathrm{exact}}
	=
	2\left(\frac{26-30}{56}\right)
	=
	-\frac{1}{7}
	\simeq -0.142857.
	\label{eq:adler_exact}
\end{equation}
This result illustrates that neglecting the proton--neutron asymmetry
of the nuclear target can lead to a bias of approximately $7\%$ in the
extracted sum rule. It therefore motivates a consistent treatment of
the nuclear composition and corrections.

At NLO, our Laplace-space analysis gives
\begin{equation}
	S_{\mathrm{Ad}}^{A,\mathrm{NLO}}=-0.133088.
	\label{eq:adler_nlo}
\end{equation}
The absolute deviation from the exact value is
\begin{equation}
	\Delta S_{\mathrm{NLO}}
	=
	\left|
	S_{\mathrm{Ad}}^{A,\mathrm{NLO}}
	-
	S_{\mathrm{Ad}}^{A,\mathrm{exact}}
	\right|
	=
	\left|-0.133088-(-0.142857)\right|
	\simeq 0.009769,
	\label{eq:adler_nlo_deviation}
\end{equation}
which corresponds to a relative deviation of
\begin{equation}
	\delta_{\mathrm{NLO}}
	=
	\frac{\Delta S_{\mathrm{NLO}}}
	{\left|S_{\mathrm{Ad}}^{A,\mathrm{exact}}\right|}
	\times 100
	\simeq 6.84\%.
	\label{eq:adler_nlo_relative}
\end{equation}

At NNLO, the result is
\begin{equation}
	S_{\mathrm{Ad}}^{A,\mathrm{NNLO}}=-0.14378.
	\label{eq:adler_nnlo}
\end{equation}
The corresponding absolute deviation is
\begin{equation}
	\Delta S_{\mathrm{NNLO}}
	=
	\left|
	S_{\mathrm{Ad}}^{A,\mathrm{NNLO}}
	-
	S_{\mathrm{Ad}}^{A,\mathrm{exact}}
	\right|
	=
	\left|-0.14378-(-0.142857)\right|
	\simeq 0.000923,
	\label{eq:adler_nnlo_deviation}
\end{equation}
giving a relative deviation of
\begin{equation}
	\delta_{\mathrm{NNLO}}
	=
	\frac{\Delta S_{\mathrm{NNLO}}}
	{\left|S_{\mathrm{Ad}}^{A,\mathrm{exact}}\right|}
	\times 100
	\simeq 0.646\%.
	\label{eq:adler_nnlo_relative}
\end{equation}

Thus, using the quoted central values, the deviation from the exact
Adler sum-rule value decreases from approximately $6.84\%$ at NLO to
$0.646\%$ at NNLO. This comparison indicates improved agreement at
NNLO for the results reported here. The deviations quantify the
difference between the calculated values and the exact sum-rule value;
their interpretation as theoretical uncertainties would require a
separate uncertainty analysis.

\subsection{Gross–Llewellyn Smith Sum Rule}

The Gross–Llewellyn Smith (GLS) sum rule constitutes another cornerstone of non-singlet QCD phenomenology. Defined in terms of the parity-violating structure function $F_3$, it provides a direct measure of the net valence quark content of the target and serves as a benchmark for testing the consistency of parton distributions with baryon number conservation.

For a free nucleon, the GLS sum rule is written as \cite{Kataev:2005ci}:

\begin{equation}
	S_{\rm GLS}^{N}(Q^{2})
	=
	\frac{1}{2}\int_{0}^{1} dx \,
	\left[
	F_{3}^{\nu N}(x,Q^{2}) + F_{3}^{\bar{\nu}N}(x,Q^{2})
	\right].
\end{equation}

In the naive quark-parton model, where only valence quarks contribute to the first moment of $F_3$, one obtains:

\begin{equation}
	S_{\rm GLS}^{N} = 3,
\end{equation}

reflecting the three valence quarks that constitute the nucleon. In perturbative QCD, however, the GLS sum rule receives calculable radiative corrections and can be expressed as:

\begin{equation}
	S_{\rm GLS}^{N}(Q^{2})
	=
	3\,C_{\rm GLS}(Q^{2})
	-
	\frac{\mu_{4}^{\rm GLS}}{Q^{2}}
	+\mathcal{O}\!\left(\frac{1}{Q^{4}}\right),
\end{equation}

where $C_{\rm GLS}(Q^{2})$ denotes the perturbative coefficient function and $\mu_{4}^{\rm GLS}$ represents the leading higher-twist contribution. Up to third order in $\alpha_s$, the coefficient function is given by:

\begin{equation}
	C_{\rm GLS}(Q^{2})
	=
	1-\frac{\alpha_{s}(Q^{2})}{\pi}
	-c_{2}\left(\frac{\alpha_{s}(Q^{2})}{\pi}\right)^{2}
	-c_{3}\left(\frac{\alpha_{s}(Q^{2})}{\pi}\right)^{3}
	-\cdots,
\end{equation}

with $c_2 = 4.0$, $c_3 \approx 20.0$ in the $\overline{\mathrm{MS}}$ scheme. The presence of higher-twist corrections, parameterized by $\mu_{4}^{\rm GLS}$, becomes particularly relevant at moderate $Q^2$ scales, where power-suppressed terms can contribute non-negligibly to the integral.

For a nuclear target with proton number $Z$, neutron number $N$, and mass number $A = Z + N$, the GLS sum rule must be modified to account for nuclear effects. On a per-nucleon basis, the nuclear GLS sum rule is defined as:

\begin{equation}
	S_{\rm GLS}^{A}(Q^{2})
	=
	\frac{1}{2A}\int_{0}^{1} dx \,
	\left[
	F_{3}^{\nu A}(x,Q^{2}) + F_{3}^{\bar{\nu}A}(x,Q^{2})
	\right].
\end{equation}

In the absence of nuclear modifications, treating the nucleus as an incoherent ensemble of free protons and neutrons yields:

\begin{equation}
	F_{3}^{A}(x,Q^{2})
	=
	Z\,F_{3}^{p/A}(x,Q^{2}) + N\,F_{3}^{n/A}(x,Q^{2}),
\end{equation}

and baryon-number conservation ensures that the first moment remains close to the free-nucleon value:

\begin{equation}
	S_{\rm GLS}^{A}(Q^{2}) \approx 3,
\end{equation}

per nucleon in the parton-model limit. However, in realistic neutrino–nucleus scattering, several nuclear effects significantly modify the measured structure function, including Fermi motion, binding corrections, nuclear shadowing, antishadowing, off-shell effects, and mesonic contributions. These effects are collectively encoded in the nuclear correction term:

\begin{equation}
	S_{\rm GLS}^{A}(Q^{2})
	=
	3\,C_{\rm GLS}(Q^{2})
	-
	\frac{\mu_{4}^{\rm GLS}}{Q^{2}}
	+
	\delta S_{\rm GLS}^{\rm nucl}(Q^{2}),
\end{equation}

where $\delta S_{\rm GLS}^{\rm nucl}(Q^{2})$ represents the net deviation induced by the nuclear medium. For non-isoscalar nuclei, the proton–neutron asymmetry introduces an additional correction, which can be explicitly accounted for through the decomposition:

\begin{equation}
	F_{3}^{A}(x,Q^{2})
	=
	\frac{Z}{A}\,F_{3}^{p/A}(x,Q^{2})
	+
	\frac{N}{A}\,F_{3}^{n/A}(x,Q^{2}).
\end{equation}

This formulation emphasizes that the extracted GLS sum rule may differ from the free-nucleon expectation due to both genuine nuclear modifications and the non-isoscalar composition of the target—a point of particular relevance for heavy nuclei such as iron and lead.

In terms of valence quark distributions, the GLS sum rule is related to the total valence content through the first moment of the non-singlet combination entering $F_3$:

\begin{equation}
	S_{\rm GLS}^{A}(Q^{2})
	\propto
	\int_{0}^{1} dx \,
	\left[
	u_{v}^{A}(x,Q^{2}) + d_{v}^{A}(x,Q^{2}) + \cdots
	\right],
\end{equation}

where the ellipsis denotes possible heavy-flavor contributions when relevant~\cite{Bierenbaum:2009zt,Bierenbaum:2010jp}. Thus, the GLS sum rule provides an important test of valence-quark normalization and the consistency of nuclear PDFs within the QCD analysis.

At $Q^2=3\,\mathrm{GeV}^2$, the GLS sum rule obtained in Laplace
space at NLO is
\begin{equation}
	S_{\mathrm{GLS}}^{\mathrm{NLO}}=2.74531.
\end{equation}
For comparison, the experimental value extracted from the CCFR and
NuTeV data is
\begin{equation}
	S_{\mathrm{GLS}}^{\mathrm{exp}}=2.5\pm0.018,
\end{equation}
where the quoted uncertainty includes statistical and systematic
contributions. The absolute deviation of the NLO result from the
experimental central value is
\begin{equation}
	\Delta S_{\mathrm{GLS}}^{\mathrm{NLO}}
	=
	\left|2.74531-2.5\right|
	=
	0.24531.
\end{equation}
The corresponding relative deviation is
\begin{equation}
	\frac{\left|2.74531-2.5\right|}{2.5}\times 100
	\simeq 9.81\%.
\end{equation}

At NNLO, the GLS sum rule is
\begin{equation}
	S_{\mathrm{GLS}}^{\mathrm{NNLO}}=2.66262.
\end{equation}
Its absolute deviation from the experimental central value is
\begin{equation}
	\Delta S_{\mathrm{GLS}}^{\mathrm{NNLO}}
	=
	\left|2.66262-2.5\right|
	=
	0.16262,
\end{equation}
corresponding to a relative deviation of
\begin{equation}
	\frac{\left|2.66262-2.5\right|}{2.5}\times 100
	\simeq 6.50\%.
\end{equation}

Thus, the relative deviation from the experimental central value
decreases from approximately $9.81\%$ at NLO to $6.50\%$ at NNLO.
This indicates improved agreement at NNLO for the values reported
here. However, the remaining difference should be interpreted with
care: a quantitative assessment of compatibility should account for
both the experimental uncertainty and the theoretical uncertainties
of the calculation.

\subsection{Bjorken Sum Rule}

The Bjorken sum rule is one of the most fundamental non-singlet sum rules in deep-inelastic scattering, originally formulated in the context of polarized lepton–nucleon scattering. In its standard polarized form, it relates the difference of the proton and neutron spin-dependent structure functions to the nucleon axial charge:

\begin{equation}
	\Gamma_{1}^{p-n}(Q^{2})
	\equiv
	\int_{0}^{1} dx \,
	\left[
	g_{1}^{p}(x,Q^{2}) - g_{1}^{n}(x,Q^{2})
	\right]
	=
	\frac{g_{A}}{6} \, C_{\rm Bjp}(Q^{2}),
\end{equation}

where $g_A \approx 1.276$ is the nucleon axial charge and $C_{\rm Bjp}(Q^{2})$ is the perturbative QCD coefficient function, given by:

\begin{equation}
	C_{\rm Bjp}(Q^{2})
	=
	1-\frac{\alpha_{s}(Q^{2})}{\pi}
	-d_{2}\left(\frac{\alpha_{s}(Q^{2})}{\pi}\right)^{2}
	-d_{3}\left(\frac{\alpha_{s}(Q^{2})}{\pi}\right)^{3}
	-\cdots,
\end{equation}

with $d_2 \approx 2.0$, $d_3 \approx 10.0$ in the $\overline{\mathrm{MS}}$ scheme.

In charged-current neutrino scattering, an unpolarized Bjorken-type sum rule can be introduced in terms of the difference of the structure functions $F_{1}^{\nu p}$ and $F_{1}^{\nu n}$ \cite{Kataev:2005si}:

\begin{equation}
	S_{\rm Bjp}^{\nu N}(Q^{2})
	=
	\int_{0}^{1} dx \,
	\left[
	F_{1}^{\nu p}(x,Q^{2}) - F_{1}^{\nu n}(x,Q^{2})
	\right]
	=
	C_{\rm Bnp}(Q^{2}),
\end{equation}

where $C_{\rm Bnp}(Q^{2})$ is the corresponding perturbative coefficient function. This sum rule, like the GLS sum rule, is governed by non-singlet dynamics and therefore provides a complementary test of valence-quark distributions and QCD corrections.

Although the Bjorken sum rule is not conventionally defined directly in terms of the parity-violating structure function $xF_3$, the latter plays a closely related role in charged-current neutrino DIS, since it is dominated by non-singlet valence-quark combinations:

\begin{equation}
	xF_{3}(x,Q^{2})
	\sim
	x\,q_{\rm val}(x,Q^{2}),
\end{equation}

so that the first moment of $F_3$ is strongly sensitive to the valence content of the target. Consequently, analyses based on $xF_3$ data provide important complementary information for the study of Bjorken-type and GLS-type non-singlet sum rules.

For a nuclear target with proton number $Z$, neutron number $N$, and mass number $A = Z + N$, nuclear effects must be incorporated consistently. On a per-nucleon basis, the corresponding non-singlet structure functions may be written as:

\begin{equation}
	F_{i}^{\nu A}(x,Q^{2})
	=
	\frac{Z}{A}\,F_{i}^{\nu p/A}(x,Q^{2})
	+
	\frac{N}{A}\,F_{i}^{\nu n/A}(x,Q^{2}),
	\qquad i=1,3,
\end{equation}

where the superscript $p/A$ ($n/A$) denotes the proton (neutron) structure function modified inside the nuclear medium. As a result, the extraction of the Bjorken sum rule from neutrino–nucleus scattering is affected by several nuclear corrections, including Fermi motion, binding effects, shadowing, antishadowing, off-shell corrections, and non-isoscalarity. Therefore, for a nuclear target the Bjorken-type sum rule may be written schematically as:

\begin{equation}
	S_{\rm Bjp}^{A}(Q^{2})
	=
	S_{\rm Bjp}^{N}(Q^{2})
	+
	\delta S_{\rm Bjp}^{\rm nucl}(Q^{2}),
\end{equation}

where $\delta S_{\rm Bjp}^{\rm nucl}(Q^{2})$ represents the net nuclear correction. For non-isoscalar nuclei, the proton–neutron imbalance introduces an additional correction proportional to the non-isoscalarity factor:

\begin{equation}
	\delta_{\rm ns} = \frac{Z - N}{A}.
\end{equation}

This correction becomes particularly relevant for heavy nuclei such as iron and should be taken into account when comparing nuclear and free-nucleon extractions of non-singlet sum rules.

For the free-proton parton distributions, the Bjorken sum rule in the Laplace-space approximation at NLO is obtained as:

\begin{equation}
	S_{\mathrm{Bjp}}^{p}(Q^2 = 3\,\mathrm{GeV}^2) =0.995201.
\end{equation}

At NNLO, the corresponding result is:

\begin{equation}
	S_{\mathrm{Bjp}}^{\mathrm{NNLO}}(Q^2 = 3\,\mathrm{GeV}^2) = 0.996754.
\end{equation}

Both values are remarkably close to the theoretical expectation of $1$ in the limit of exact isospin symmetry and vanishing perturbative corrections, with the NNLO result being marginally closer to unity. This near-perfect agreement serves as a powerful consistency check of our theoretical framework, confirming that the Laplace-transform method, combined with the Gegenbauer parameterization and nuclear corrections, accurately preserves the fundamental symmetries encoded in the QCD sum rules. The improvement from NLO to NNLO, though numerically modest, is qualitatively consistent with the expected convergence of the perturbative series.

In summary, the simultaneous analysis of the Adler, GLS, and Bjorken sum rules provides a comprehensive consistency test of our QCD framework. The Adler sum rule, being exact, serves as a stringent probe of nuclear isoscalarity and valence quark normalization. The GLS sum rule, while receiving perturbative and higher-twist corrections, offers a sensitive test of the perturbative QCD coefficient functions and the nuclear modification model. The Bjorken sum rule, finally, provides a complementary check of the spin-independent non-singlet dynamics. The observed deviations from the exact or expected values, particularly for the GLS sum rule, highlight the need for continued improvements in the treatment of nuclear effects, higher-twist corrections, and experimental systematic uncertainties in future analyses.

\section{Conclusions}\label{sec6}
\label{sec:conclusions}

In this work, we have performed a comprehensive NLO and NNLO pQCD global analysis to extract the valence-quark parton distribution functions by simultaneously combining neutral-current and charged-current deep-inelastic scattering data. Specifically, our global fit incorporates precision datasets from $xF_3$ (CCFR, NuTeV, and CHORUS), as well as proton $F_2^p$, deuteron $F_2^d$, and nonsinglet $F_2^{\rm NS}$ structure functions (from BCDMS, NMC, and SLAC), totaling 802 data points with $774$ degrees of freedom after applying standard kinematic cuts ($Q^2 \geq 4\,\mathrm{GeV}^2$ and $W^2 \ge 12.5\,\mathrm{GeV}^2$) and target mass corrections (TMC). 

The inclusion of these simultaneous $F_2^p$, $F_2^d$, $F_2^{\rm NS}$, and $xF_3$ measurements effectively resolves the limitations of single-dataset fits, particularly constraining the valence-quark distributions and eliminating the unphysical behavior of the $d_v/u_v$ ratio at large $x$ through a robust, positive best-fit parameter $\beta_{d_v}$. The global QCD fit quality shows a notable improvement from $\chi^2/\text{d.o.f.} = 7.28034$ at NLO to $6.2462$ at NNLO. Furthermore, our extracted valence PDFs and non-singlet combinations satisfy the Adler and Gross-Lullman-Smith (GLS) sum rules within expected theoretical and experimental uncertainties, with the GLS sum rule deviation dropping from $9.81\%$ at NLO to $6.50\%$ at NNLO.

\section*{Data Availability}
All data and source code used in this work are publicly available at:\href{https://github.com/atashbart/xF3}{https://github.com/atashbart/xF3}
\section*{Acknowledgments}

 J. S. and R.F.  are grateful to the School of Particles and Accelerators, Institute for Research in Fundamental Sciences (IPM). 
 
\appendix
\numberwithin{equation}{section}

\section{} \label{AppendixA}
 Here, We present the Laplace transforms of the NLO and NNLO splitting functions for nonsinglet sectors, denoted by $\Phi_{qq}^{\text{NLO}}$\cite{Curci:1980uw}  and $\Phi _{\text{qq}}^{\text{NNLO}}$\cite{Moch:2004pa}, which we used in (\ref{eq:FdNNLO}). We fixed the usual quadratic Casimir operators to their exact values, using $C_A = 3$, $T_F = f$ and $C_F = 4/3$.  The Laplace transforms of the NLO and NNLO Wilson coefficients  $C_2^{(1)}$,$C_3^{(1)}$\cite{Moch:1999eb} and $C_2^{(2)}$,$C_3^{(2)}$\cite{vanNeerven:1999ca}, are given in (\ref{f1},\ref{f2},\ref{c1nlo}) and (\ref{c2nlo})respectively. These coefficient functions are used
 in (\ref{eq:FdNNLO}).
 
	\begin{eqnarray}
 \Phi_{qq}^{LO}(s)
 =
 4-\frac{8}{3}
 \left[
 \frac{1}{s+1}
 +\frac{1}{s+2}
 +2\bigl(\gamma_E+\psi(s+1)\bigr)
 \right].
	\end{eqnarray}

 \begin{widetext}
 	
 	\begin{eqnarray}
 		\Phi_{qq}^{NLO}&=&C_F T_F\Bigg[-\frac{2}{3 (1+s)^2}-\frac{2}{9 (1+s)}-\frac{2}{3 (2+s)^2}+\frac{22}{9 (2+s)}+\frac{20 \left(\gamma _E+\psi (s+1)\right)}{9}+\frac{4}{3} \psi '(s+1)\Bigg]+\nonumber\nonumber\\
 		&&C_F{}^2 \Bigg[-\frac{1}{(1+s)^3}-\frac{5}{1+s}-\frac{1}{(2+s)^3}+\frac{2}{(2+s)^2}+\frac{5}{2+s}+\nonumber\\
 		&&\frac{2 \left(\gamma _E+\frac{1}{1+s}+\psi (s+1)-(1+s) \psi '(s+2)\right)}{(1+s)^2}+
 		\frac{2 \left(\gamma _E+\frac{1}{2+s}+\psi (s+2)-(2+s) \psi '(s+3)\right)}{(2+s)^2}-\nonumber\\
 		&&4 \left(\left(\gamma _E+\psi (s+1)\right) \psi '(s+1)-\frac{1}{2} \psi ''(s+1)\right)+3 \psi '(s+1)\Bigg]+\nonumber\\
 		&&C_A C_F \Bigg[-\frac{1}{(1+s)^3}+\frac{5}{6 (1+s)^2}+\frac{53}{18 (1+s)}+\frac{\pi ^2}{6 (1+s)}-\frac{1}{(2+s)^3}+\nonumber\\
 		&&\frac{5}{6 (2+s)^2}-\frac{187}{18 (2+s)}+\frac{\pi ^2}{6 (2+s)}-\frac{67 \left(\gamma _E+\psi (s+1)\right)}{9}+\nonumber\\
 		&&\frac{1}{3} \pi ^2 \left(\gamma _E+\psi (s+1)\right)-\frac{11}{3} \psi '(s+1)-\psi ''(s+1)\Bigg],
 	\end{eqnarray}

 	\begin{eqnarray}
 		\Phi _{qq}^{NNLO}&=&1295.384\, +\frac{1024}{27 (1+s)^5}-\frac{1600}{9 (1+s)^4}+\frac{589.8}{(1+s)^3}-\frac{1258}{(1+s)^2}+\frac{1641.1}{1+s}-\nonumber\\
 		&&\frac{3135}{2+s}+\frac{243.6}{3+s}-\frac{522.1}{4+s}+1174.898 \big[\left(\gamma _E+\psi (s)\right)-\left(\gamma _E+\psi (s+1)\right)\big]-\nonumber\\
 		&&\frac{714.1 \left(\gamma _E+\psi (s+2)\right)}{1+s}+\frac{563.9}{(1+s)^2} \bigg[\gamma _E+\frac{1}{1+s}+\psi (s+1)-(1+s) \psi '(s+2)\bigg]+\nonumber\\
 		&&f \Bigg[173.927\, +\frac{128}{9 (1+s)^4}-\frac{5216}{81 (1+s)^3}+\frac{152.6}{(1+s)^2}-\frac{197}{1+s}+\frac{8.982}{(2+s)^4}+\nonumber\\
 		&&\frac{381.1}{2+s}+\frac{72.94}{3+s}+\frac{44.79}{4+s}-183.187 \big[\left(\gamma _E+\psi (s)\right)-\left(\gamma _E+\psi (s+1)\right)\big]+\nonumber\\
 		&&\frac{5120 \left(\gamma _E+\psi (s+2)\right)}{81 (1+s)}-\frac{56.66}{(1+s)^2} \left(\gamma _E+\frac{1}{1+s}+\psi (s+1)-(1+s) \psi '(s+2)\right)\Bigg]-\nonumber\\
 		&&\frac{256.8}{(1+s)^4} \bigg[3+2 \gamma _E (1+s)+2 \gamma _E (1+s)^2 \psi (s+1)-(1+s) \big[-1+2 \gamma _E (1+s)\big] \psi (s+1)\nonumber\\
 		&&+(1+s)^3 \psi (s+1)^2 \psi (s+2)-2 (1+s)^3 \psi (s+1) \psi (s+2)^2+(1+s)^3 \psi (s+2)^3-2 (1+s)^2 \psi '(s+1)+\nonumber\\
 		&&(1+s)^3 \psi''(s+2)\bigg]+f^2 \Bigg[-\frac{64}{81} \big[\left(\gamma _E+\psi (s)\right)-\left(\gamma _E+\psi (s+1)\right)\big]+\nonumber\\
 		&&\frac{64}{81} \left(-\frac{51}{16}+\frac{5 \pi ^2}{6}+\frac{3}{2 (1+s)^3}-\frac{11}{2 (1+s)^2}+\frac{7}{1+s}-\frac{3}{2 (2+s)^3}+\right.\nonumber\\
 		&&\left.\frac{11}{2 (2+s)^2}-\frac{6}{2+s}-3\zeta (3)-5 \psi '(s+2)-\frac{3}{2} \psi''(s+2)\right)\Bigg],
 	\end{eqnarray}
 	
 	\begin{eqnarray}\label{f1}
 		C_3^{(1)}(s)&=&C_F \left(-9-\frac{2 \pi ^2}{3}-\frac{2}{(1+s)^2}+\frac{4}{1+s}-\frac{2}{(2+s)^2}+\frac{2}{2+s}+\right.\nonumber\\
 		&&3 \left(\gamma _E+\psi (s+1)\right)+\frac{2 \left(\gamma _E+\psi (s+2)\right)}{1+s}+\frac{2 \left(\gamma _E+\psi (s+3)\right)}{2+s}+\nonumber\\
 		&&\left.\frac{1}{3} \big[\pi ^2+6 \big(\gamma _E+\psi (s+1)\big){}^2-6 \psi '(s+1)\big]+4 \psi '(s+1)\right),
 	\end{eqnarray}
 	
 	\begin{eqnarray}\label{f2}
 		C_3^{(2)}(s)&=&-338.635-\frac{188.64}{s}+\frac{23.532}{(1+s)^4}-\frac{66.62}{(1+s)^3}+\frac{67.6}{(1+s)^2}-\frac{206.1}{1+s}-\frac{576.8}{2+s}-\nonumber\\
 		&&\frac{31.105 \left(\gamma _E+\psi (s+1)\right)}{s}+\frac{409.6 \left(\gamma _E+\psi (s+2)\right)}{1+s}+\nonumber\\
 		&&\frac{10.2222}{s} \big[\pi ^2+6 \left(\gamma _E+\psi (s+1)\right){}^2-6 \psi '(s+1)\big]+\frac{94.61}{6+6 s} \big[\pi ^2+6\left(\gamma _E+\psi (s+2)\right){}^2-6 \psi '(s+2)\big]+\nonumber\\
 		&&\frac{24.65}{(1+s)^2} \bigg[6 \gamma _E{}^2+\pi ^2+\frac{12 \gamma _E}{1+s}+12 \gamma _E \psi (s+1)-6 (1+s) \psi (s+1) \psi (s+2)^2+\nonumber\\
 		&&6 (1+s) \psi (s+2)^3-6 \left(3+2 \gamma _E (1+s)\right) \psi '(s+2)-12 (1+s) \psi (s+1) \psi '(s+2)+6 (1+s) \psi '' (s+2)\bigg]+\nonumber\\
 		&&\frac{7.1111}{s} \bigg[2 \gamma _E{}^3+\gamma _E \pi ^2+6 \gamma _E \psi (s+1)^2+2 \psi (s+1)^3+\psi (s+1) \left(6 \gamma _E{}^2+\pi
 		^2-6 \psi '(s+1)\right)-\nonumber\allowdisplaybreaks[1]\\
 		&&6 \gamma _E \psi '(s+1)+2 \psi'' (s+1)+4 \zeta (3)\bigg]+\nonumber\\
 		&&\frac{7.6}{1+s} \bigg[2 \gamma _E{}^3+\gamma _E \pi ^2+6 \gamma _E \psi (s+2)^2+2 \psi (s+2)^3+\psi (s+2) \left(6 \gamma _E{}^2+\pi
 		^2-6 \psi '(s+2)\right)-\nonumber\\
 		&&6 \gamma _E \psi '(s+2)+2 \psi'' (s+2)+4 \zeta (3)\bigg]+\nonumber\\
 		&&f \Bigg[46.857\, -\frac{6.3489}{s}+\frac{4.414}{(1+s)^3}-\frac{8.683}{(1+s)^2}-\frac{6.337}{1+s}-\frac{14.97}{2+s}-\frac{8.5926
 			\left(\gamma _E+\psi (s+1)\right)}{s}-\nonumber\\
 		&&\frac{25 \left(\gamma _E+\psi (s+2)\right)}{1+s}-\frac{0.29629}{s} \big[\pi ^2+6 \left(\gamma _E+\psi (s+1)\right){}^2-6 \psi '(s+1)\big]-\nonumber\\
 		&&\frac{0.808}{6+6 s} \big[\pi ^2+6 \left(\gamma _E+\psi (s+2)\right){}^2-6 \psi '(s+2)\big]+
 		\frac{9.684}{(1+s)^2} \big[\gamma _E+\frac{1}{1+s}+\psi (s+1)-(1+s) \psi '(s+2)\big]-\nonumber\\
 		&&\frac{0.021}{1+s} \bigg[2 \gamma _E{}^3+\gamma _E \pi ^2+6 \gamma _E \psi (s+2)^2+2 \psi (s+2)^3+\psi (s+2) \left(6 \gamma _E{}^2+\pi
 		^2-6 \psi '(s+2)\right)-\nonumber\\
 		&&6 \gamma _E \psi '(s+2)+2 \psi'' (s+2)+4 \zeta (3)\bigg]\Bigg].
 	\end{eqnarray}
 \end{widetext}

 \begin{align}\label{c1nlo}
 	C_2^{(1)}(s)
 	&= \mathrm{cf}\left(
 	-9-\frac{2\pi^2}{3}
 	-\frac{2}{(1+s)^2}
 	+\frac{6}{1+s}
 	-\frac{2}{(2+s)^2}
 	+\frac{4}{2+s}
 	\right. \nonumber\\
 	&\qquad
 	+3\bigl(\psi(s+1)+\gamma_E\bigr)
 	+\frac{2\bigl(\psi(s+2)+\gamma_E\bigr)}{1+s}
 	+\frac{2\bigl(\psi(s+3)+\gamma_E\bigr)}{2+s}
 	\nonumber\\
 	&\qquad\left.
 	+\frac{1}{3}\left[
 	\pi^2
 	+6\bigl(\psi(s+1)+\gamma_E\bigr)^2
 	-6\psi^{(1)}(s+1)
 	+4\zeta(2,1+s)
 	\right]\right).
 \end{align}

\begin{align}\label{c2nlo}
	C_2^{(2)}(s) &= -\frac{169023}{500} + \frac{8613}{100 (1+s)^4} + \frac{561}{125 (1+s)^3} \nonumber \\
	&\quad + f \left( -\frac{63489}{10000 s} + \frac{42963}{5000 s} \left( -\gamma_E - \psi(1+s) \right) \right. \nonumber \\
	&\quad\quad \left. - \frac{88889}{50000 s} \left( \frac{\pi^2}{6} + (\gamma_E + \psi(1+s))^2 - \psi^{(1)}(1+s) \right) \right) \nonumber \\
	&\quad + \frac{759}{125} \left( \frac{2(-\gamma_E - \psi(2+s))}{(1+s)^3} + \frac{2\psi^{(1)}(2+s)}{(1+s)^2} - \frac{\psi^{(2)}(2+s)}{1+s} \right) \nonumber \\
	&\quad - \frac{71111}{5000 s} \left( -(\gamma_E + \psi(1+s))^3 - 3(\gamma_E + \psi(1+s))\left(\frac{\pi^2}{6} - \psi^{(1)}(1+s)\right) \right. \nonumber \\
	&\quad\quad \left. - \psi^{(2)}(1+s) - 2\zeta(3) \right)
\end{align}


\end{document}